\documentclass[12pt,a4]{article}

\usepackage{amsmath}
\usepackage{amssymb}
\usepackage{graphicx}
\usepackage{subfigure}
\usepackage{cite}
\usepackage{alltt}

\begin{document}

\begin{titlepage}

\vspace{1mm}
\begin{center}{\bf\Large {THERMINATOR}: \\\vspace{2mm}
THERMal heavy-IoN generATOR$^{\dag}$}
\end{center}

\vspace{3 mm}

\begin{center}
{\large  Adam Kisiel$^{\,a}$,}
{\large  Tomasz Ta\l{}u\'c$^{\,a}$,}
{\large  Wojciech Broniowski$^{\,b}$, and} 
{\large  Wojciech Florkowski$^{\,c,b}$}

\vspace{0.5cm} 
{\em $^a$  Faculty of Physics, Warsaw University of Technology,
PL-00661 Warsaw, Poland}\\ 
{\em $^b$ The H. Niewodnicza\'nski Institute of
Nuclear Physics, \\ Polish Academy of Sciences, PL-31342 Cracow,
Poland}\\ 
{\em $^c$ Institute of Physics, \'Swi\c{e}tokrzyska Academy,
PL-25406 Kielce, Poland}
\end{center}
\vspace{1mm}
\begin{abstract}
{\tt THERMINATOR} is a Monte Carlo event generator designed for studying of 
particle production in relativistic heavy-ion collisions performed at such
experimental facilities as the SPS, RHIC, or LHC. The program implements   
{\em thermal models} of particle production with {\em single freeze-out}.
It performs the following tasks: 1)~generation of stable particles and unstable 
resonances at the chosen freeze-out hypersurface with the local phase-space density 
of particles given by the statistical distribution factors, 2)~subsequent 
space-time evolution and decays of hadronic resonances in cascades,  3)~calculation 
of the transverse-momentum spectra and numerous other observables related to the 
space-time evolution. The geometry of the freeze-out hypersurface 
and the collective velocity of expansion may be chosen from two successful models, 
the Cracow single-freeze-out model and the Blast-Wave model. All particles from 
the Particle  Data Tables are used. 
The code is written in the object-oriented {\tt c++} language and 
complies to the standards of the {\tt ROOT} environment. 
\end{abstract}

\begin{center}
{\it Submitted to Computer Physics Communications}
\end{center}

\vspace{1mm}
\footnoterule
\noindent
{\footnotesize
\begin{itemize}
\item[${\dag}$] Work supported in part by 
the Polish State Committee for Scientific Research grant 2~P03B 059 25. 
\end{itemize}
}

\end{titlepage}



\noindent{\bf PROGRAM SUMMARY}
\vspace{0.3cm}

\noindent{\sl Manuscript title:}  Therminator - THERMal heavy IoN generATOR
\vspace{0.3cm}

\noindent{\sl Authors:} \- Adam Kisiel, Tomasz Ta\l{}u\'c, Wojciech Broniowski, Wojciech Florkowski
\vspace{0.3cm}

\noindent{\sl Program title:} \- {\tt THERMINATOR}, \hfill April 2005, version 1.0;
\vspace{0.3cm}

\noindent{\sl RAM required to execute with typical data:} 50 Mbytes
\vspace{0.3cm}

\noindent{\sl Number of processors used:} 1
\vspace{0.3cm}

\noindent{\sl Computer(s) for which the program has been designed:}\\
PC, Pentium III,IV, or Athlon, 512~MB RAM \hfill \lowercase{NOT HARDWARE DEPENDENT};
(any computer with the {\tt c++} compiler and the {\tt ROOT} environment \cite{root})
\vspace{0.3cm}

\noindent{\sl Operating system(s) for which the program has been designed:}\\
{\bf Linux:} Mandrake 9.0, Debian 3.0, SuSE 9.0, Red Hat FEDORA~3, {\em etc.},  
{\bf Windows XP} with Cygwin ver.~1.5.13-1 and {\tt gcc} ver.~3.3.3
(cygwin special);  \hfill {-- NOT  SYSTEM DEPENDENT}  
\vspace{0.3cm}

\noindent{\sl External routines/libraries used:}\\
 ROOT ver. 4.02.00;
\vspace{0.3cm}

\noindent{\sl Programming language:}\-
{\tt c++} 
\vspace{0.3cm}

\noindent{\sl Size of the package:}\- 
(324  KB  directory 
40 KB compressed distribution archive), without the {\tt ROOT} libraries (see
http://root.cern.ch 
for details on the ROOT \cite{root} requirements).
The output files created by the code need 1.1 GB for each 500 events. 
\vspace{0.3cm}

\noindent{\sl Distribution format:}\- 
tar gzip file
\vspace{0.3cm}

\noindent{\sl Number of lines in distributed program, including test data, etc.:}\- 
6516
\vspace{0.3cm}

\vspace{0.3cm}

\vspace{0.3cm}

\noindent{\sl Nature of the physical problem:}\\
Statistical models have proved to be very useful in the description of  soft physics 
in relativistic heavy-ion collisions \cite{Braun-Munzinger:2003zd}. In particular,
with a few physical input parameters, such as the temperature, chemical potentials, and
velocity of the collective flow, the models reproduce the observed particle
abundances \cite{Koch:1985hk,Cleymans:1992zc,Sollfrank:1993wn,Braun-Munzinger:1994xr,
Braun-Munzinger:1995bp,Cleymans:1996cd,Becattini:1997uf,Yen:1998pa,Braun-Munzinger:1999qy,
Cleymans:1999st,Becattini:2000jw,Braun-Munzinger:2001ip,Florkowski:2001fp},
the transverse-momentum spectra \cite{Broniowski:2001we}, balance functions 
\cite{Florkowski:2004em,Bozek:2003qi}, or the elliptic flow
\cite{Broniowski:2002wp,Florkowski:2004du} in both non-strange and strange 
sectors. The key element of the approach is the inclusion of the complete list of 
hadronic resonances, 
which at the rather high temperature at freeze-out, $\sim 165$~MeV, contribute very 
significantly to the observed quantities. Their two- and three-body decays, taken from the 
tables, proceed in cascades, ultimately producing the stable particles observed in detectors.
At the moment there exist several codes to compute the abundances of particles
(the publicly available programs for this purpose are {\tt SHARE} \cite{Torrieri:2004zz} and {\tt THERMUS} 
\cite{Wheaton:2004qb}), which is a rather simple task, since the abundances are insensitive 
to the geometry of the fireball and its expansion. On the other hand, the calculation of the 
transverse-momentum spectra of particles is much more complicated due to the sensitivity to 
these phenomena. {\tt THERMINATOR} deals with this problem, offering the {\em full 
information on the space-time positions and momenta of the produced particles}. As a result, 
the program allows to compute very efficiently the transverse-momentum spectra of identified particles 
and examine implications of the assumed expansion model. {\tt THERMINATOR} allows easily for the 
departure from symmetries typically assumed in other approaches. This opens the possibility to 
study the dependence of physical quantities on rapidity and the azimuthal angle. The 
contribution of the resonances to various observables may be traced conveniently, and their 
role in the statistical approach may be verified. As a Monte Carlo event generator 
written in the object-oriented {\tt c++} language in the {\tt ROOT} \cite{root} 
environment,  {\tt THERMINATOR} can be 
straightforwardly interfaced to the standard software routinely used in the data analysis 
for relativistic heavy-ion colliders, such as SPS, RHIC, and, in the future, LHC. In
this way the inclusion of experimental acceptance, kinematic cuts, or interfacing with 
other programs poses no difficulty.

\vspace{0.3cm}

\noindent{\sl Method of solving the problem:}\\
{\tt THERMINATOR} uses the particle data tables \cite{Hagiwara:2002fs} in the universal input 
form used 
by the {\tt SHARE} \cite{Torrieri:2004zz} package. The user decides for the thermal parameters and the preferred  expansion 
model. The optimum thermal parameters may be taken, {\em e.g.},  as those obtained with the help 
of {\tt SHARE} \cite{Torrieri:2004zz} or {\tt THERMUS} \cite{Wheaton:2004qb}. At the moment there are 
two different expansion models implemented in the code: the model 
of Ref.~\cite{Broniowski:2001we}, based on the so-called Buda-Lund \cite{Csorgo:1995bi} 
parameterization, 
and the Blast-Wave model \cite{Schnedermann:1993ws,Retiere:2003kf}. 
The positions and velocities of the particles are randomly generated on the hypersurface according 
to the statistical (Bose-Einstein of Fermi-Dirac) distribution factors. All particles, 
stable and unstable, are included. The particles move along classical trajectories from their
initial positions, with velocities composed of the thermal motion and the collective expansion
of the system. Stable particles just stream freely, while the resonances decay after some 
(randomly generated) time, which is controled by the particle's lifetime. The decays are 
two-body or three-body, and their implementation involves simple kinematic formulas. The
decays can 
proceed in cascades, down to the stage where only stable particles are present. All particles 
have tags indicating their parent. The secondary rescatterings are not
considered in this approach.  Full history of the event is stored in an output file, allowing for a 
detailed examination of the space-time evolutions and the calculation of the 
transverse-momentum spectra. 

\vspace{0.3cm}

\noindent{\sl Purpose:}\\
The ongoing analyses of the SPS and the RHIC data as well as the future heavy-ion 
program at LHC will certainly benefit from {\tt THERMINATOR} as a tool for generating events in a 
simple statistical model. The Monte Carlo code written in {\tt c++} and using the standard 
{\tt ROOT} \cite{root} 
environment can be easily adapted to purposes directly linked to experimental data analyses. 
The space-time tracking capability will allow, in the framework of the statistical
approach, to better understand the physics of relativistic heavy-ion collisions.
{\tt THERMINATOR} calculates the particle spectra and other observables
related to the space-time evolution of the system. It provides a {\tt c++} framework which may be easily
developed for detailed analyses of more involved 
observables such as, {\em e.g.}, correlation functions or HBT radii.

\vspace{0.3cm}

\noindent{\sl Computation time survey and storage requirements:} \\
The generation of 500 events from scratch takes about 1 hour 15 minutes
on a PC with Athlon-Barthon 2.5~GHz under Red Hat Fedora 3. Each subsequent 500 events 
take about 1 hour. To store 500 events about 1.1 GB disk storage is needed, depending on
the kinematic range. After converting the output to the ROOT TTree format, 900 MB may be freed. 

\vspace{0.3cm}

\noindent{\sl Accessibility:} \\
{\tt http://hirg.if.pw.edu.pl/en/therminator} \\

\section{Introduction}

Statistical description of heavy-ion experiments has proven to be very successful
in understanding and quantitatively describing the {\em soft} physics, {\em i.e.},  the 
regime with transverse-momentum scales not exceeding, say, 2~GeV. The simple basic assumption 
that the hadronic matter reaches thermal equilibrium and undergoes rapid expansion 
\cite{Fermi:1950jd,Pomeranchuk:1951ey,Landau:1953gs,Hagedorn:1965st} 
leads to a remarkably good description of the ratios of 
particle abundances measured in heavy-ion experiments, in particular at the high 
energies of the SPS ($\sqrt{s_{NN}}$=8--17~GeV) and
RHIC (reaching $\sqrt{s_{NN}}$=200~GeV). The results for abundances may be found in 
\cite{Koch:1985hk,Cleymans:1992zc,Sollfrank:1993wn,Braun-Munzinger:1994xr,
Braun-Munzinger:1995bp,Cleymans:1996cd,Becattini:1997uf,Yen:1998pa,Braun-Munzinger:1999qy,
Cleymans:1999st,Becattini:2000jw,Braun-Munzinger:2001ip,Florkowski:2001fp}.
Other details of different statistical approaches may be found in
\cite{Schnedermann:1993ws,Retiere:2003kf,Rafelski:1996hf,Rafelski:1997ab,Csorgo:1995bi,
Csorgo:1999sj,Cleymans:1998fq,Gazdzicki:1998vd,Gazdzicki:1999ej,Broniowski:2002nf}.

Crucial for the success of the statistical approach is the complete inclusion of hadronic 
resonances, whose number grows rapidly according to the Hagedorn hypothesis 
\cite{Hagedorn:1965st,Hagedorn:1968ua,Hagedorn:1994sc,Broniowski:2000bj,Broniowski:2000hd,
Broniowski:2004yh}.
Numerous programs exist to analyze the ratios of particle abundances, with {\tt SHARE} 
\cite{Torrieri:2004zz} and {\tt THERMUS} \cite{Wheaton:2004qb} publicly available. 
The abundances are observables that are sensitive to the thermal parameters 
of the fireball created in the collision: temperature, chemical potentials, and fugacity 
factors. The abundances are basically insensitive to the expansion\footnote{Some dependence 
may come for boost non-invariant systems from the detector acceptance, however in the most 
studied mid-rapidity region at RHIC and at SPS this is a tiny effect.}, which means that 
it only matters how many particles are formed and it is irrelevant how fast they move.

The situation is different for the description of the momentum spectra, where the 
collective motion of the expansion adds up to the thermal one generated in the 
local element of the fireball. Thus, the spectra are sensitive not only to the thermal
parameters, but also to the geometry of the fire-ball and velocity of expansion.
Surprisingly, very simple hypotheses for the geometry and expansion 
\cite{Broniowski:2001we} allow for a good description of the RHIC data for the 
transverse-momentum spectra of identified particles, both in the non-strange and strange 
sectors \cite{Broniowski:2001uk}. The approach of Ref.~\cite{Broniowski:2001we} is based 
on the simplifying assumption that the so-called chemical and thermal freeze-out occur 
simultaneously thus we deal with a {\em single-freeze-out model}. Similar hypothesis is 
otherwise known as the {\em sudden hadronization} \cite{Rafelski:2000by}. We follow this 
paradigm in {\tt THERMINATOR} and throughout this paper \footnote{Since the program is an event 
generator, it would not be very complicated to add an ``afterburner'' that would scatter 
the particles elastically after they have frozen chemically, thus differentiating between 
the two freeze-outs.}. 

{\tt THERMINATOR} computes the transverse-momentum spectra of identified
particles in the single-freeze-out model. The user has an option of choosing 
two variants of the geometry/expansion models: the Buda-Lund type expansion 
\cite{Csorgo:1995bi,Csorgo:1999sj} and the Blast-Wave type expansion 
\cite{Schnedermann:1993ws,Retiere:2003kf} (both models of expansion were used with a 
similar success in the original single-freeze-out model \cite{Broniowski:2001we}). Other 
expansion models may be looked at with the need of only minor modifications to the code. 
As already mentioned, the key ingredient in the thermal approach are the hadronic resonances, 
which must be included in a complete way, together with their decay channels and branching 
ratios. {\tt THERMINATOR} uses the same universal input files with the information from the 
Particle Data Tables \cite{Hagiwara:2002fs} as the {\tt SHARE} package \cite{Torrieri:2004zz} 
co-authored by two of us (WB,WF). The basic functionality of {\tt THERMINATOR} is the calculation 
of the momentum spectra and, correspondingly, the analysis of the geometry and expansion. 
It allows for the optimum determination of the transverse flow in the assumed model of 
expansion. This may be very useful for SPS, RHIC, and the future heavy-ion program 
at LHC. However, the program is a versatile Monte Carlo {\rm event generator}, and as 
such it may be used with slight modifications to analyze other observables, such as HBT 
correlation radii
\cite{Baym:1997ce,Wiedemann:1999qn,Heinz:1999rw,Csorgo:1999sj,Tomasik:2002rx}, 
correlations of non-identical particles, or balance functions.
With simple modifications amounting to a departure from the presently-implemented boost and
azimuthal symmetries, one may look at the dependence of parameters and observables on 
rapidity, or at measures of the azimuthal asymmetry, such as the coefficient of the elliptic 
flow $v_2$. We want to emphasize that {\tt THERMINATOR} uses the {\tt ROOT} environment, thus it may 
be conveniently linked to the standard software used for the data analysis.

\section{Basics of the single-freeze-out model}
\label{sect:basics}

The basic assumptions of the single-freeze-out model are as follows 
(for a detailed discussion see 
Refs.~\cite{Broniowski:2001we,Broniowski:2001uk,Broniowski:2002nf}):

\begin{enumerate}
\item At some point of the space-time evolution of the fireball the local thermal 
and chemical equilibrium
is reached. The local particle phase-space densities follow the Fermi-Dirac or Bose-Einstein 
statistical distributions. 
\item Freeze-out occurs at universal values of the thermodynamic parameters. 
We have temperature $T$ and three chemical potentials: baryon, $\mu_B$, 
strange, $\mu_S$, and related to the third component of isospin, $\mu_{I_3}$. Typically, the potentials 
$\mu_S$ and $\mu_{I_3}$ are set from the condition of the overall strangeness neutrality 
and from the assumption that the ratio of the electric to baryon charge density equals 
the initial ratio, $Z/A$, where $Z$ and $A$ are the atomic and mass numbers of the 
colliding nuclei. The values of the thermodynamic parameters are obtained from fits to the 
measured ratios of the particle abundances. The publicly-available
{\tt SHARE} \cite{Torrieri:2004zz} or {\tt THERMUS}
\cite{Wheaton:2004qb} codes may be used here. 
\item The freeze-out occurs on a boost-invariant
cylindrically symmetric hypersurface, discussed in detail below. The velocity field of 
expansion is chosen appropriately, providing the longitudinal and transverse flow to the 
system. The particles generated at freeze-out are termed {\it primordial}.
\item The later evolution consists of decays of resonances which may proceed in 
cascades. All resonances from the Particle Data Tables \cite{Hagiwara:2002fs} 
are incorporated.   
\end{enumerate}

\begin{figure}[tb]
\begin{center}
\includegraphics[width=9cm]{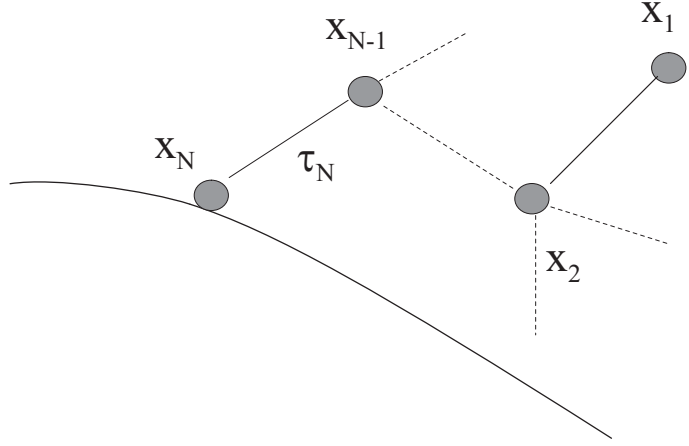}
\end{center}
\caption{The cascade decay in the single-freeze-out model. An unstable particle  
is formed at some position on the hypersurface with the momentum composed of the thermal 
motion and the flow of the medium. Then the particle travels for some time depending on 
its lifetime and decays. The products, if resonances themselves, may decay further, 
until the stable particles are formed.}
\label{soup}
\end{figure}

The basic physics of the model is depicted in Fig.~\ref{soup}. The popular choices of 
the hypersurface of freeze-out and the velocity field of expansion 
are discussed in detail in Ref.~\cite{Florkowski:2004tn}. The boost-invariant formalism 
has been developed by Schnedermann, Sollfrank, and Heinz \cite{Schnedermann:1993ws}.
The ansatz for the boost-invariant, cylindrically symmetric freeze-out hypersurface
has the form
\begin{equation}
x^\mu = (t,x,y,z) =
 \left( {\tilde \tau}(\zeta) \hbox{cosh } \alpha_\parallel,
\rho(\zeta) \cos\phi, \rho(\zeta) \sin\phi, 
{\tilde \tau}(\zeta) \hbox{sinh } \alpha_\parallel \right).
\label{xmubinv}
\end{equation}
The parameter $\alpha_\parallel$ is the space-time rapidity. 
The boost-invariant freeze-out hypersurface is completely 
defined by the freeze-out curve parameterized by $\zeta$. 
This curve defines the freeze-out times of the cylindrical shells with the radius 
$\rho$. The volume element of the freeze-out hypersurface is obtained from the 
equation
\begin{equation}
d^3\Sigma_\mu = \varepsilon_{\mu \alpha \beta \gamma}
{d x^\alpha \over d\alpha }
{d x^\beta \over d\beta }
{d x^\gamma \over d\gamma } d\alpha d\beta d\gamma,
\label{d3sigma}
\end{equation}
which with the parameterization (\ref{xmubinv}) yields
\begin{equation}
d^3\Sigma^\mu = \left( {d\rho \over d\zeta} \hbox{cosh }\alpha_\parallel,
{d {\tilde \tau} \over d\zeta} \cos\phi,
{d {\tilde \tau} \over d\zeta} \sin\phi,
{d\rho \over d\zeta} \hbox{sinh }\alpha_\parallel \right)
\rho(\zeta){\tilde \tau}(\zeta) d\zeta d\alpha_\parallel d\phi.
\label{d3sigmabinv}
\end{equation}
Similarly to Eq. (\ref{xmubinv}) the boost-invariant four-velocity field has the 
structure
\begin{equation}
u^\mu = \hbox{cosh}\,\alpha_\perp(\zeta) 
\left( \hbox{cosh}\,\alpha_\parallel , \hbox{tanh}\,\alpha_\perp(\zeta) \cos\phi,
\hbox{tanh}\,\alpha_\perp(\zeta) \sin\phi,
\hbox{sinh}\,\alpha_\parallel \right). 
\label{umubinv}
\end{equation}
We note that the longitudinal flow is $v_z = \hbox{tanh}\, \alpha_\parallel = z/t$
(as in the one-dimensional Bjorken model \cite{Bjorken:1983qr}), while the transverse
flow at $z=0$ has the form $v_r = \hbox{tanh}\,\alpha_\perp(\zeta)$.

With the standard parameterization of the particle four-momentum 
in terms of rapidity $y$, azimuthal angle $\varphi$, and the transverse mass $m_\perp$,
\begin{equation}
p^\mu = \left(m_\perp \hbox{cosh} y, p_\perp \cos\varphi, p_\perp
\sin\varphi, m_\perp \hbox{sinh} y \right),
\label{pmubinv}
\end{equation}
we find
\begin{equation}
p \cdot u = m_\perp \hbox{cosh}(\alpha_\perp) 
\hbox{cosh}(\alpha_\parallel-y) -  p_\perp  
\hbox{sinh}(\alpha_\perp) \cos(\phi-\varphi),
\label{pubinv}
\end{equation}
and
\begin{equation}
d^3\Sigma  \cdot p = \left[m_\perp \hbox{cosh}(y-\alpha_\parallel) 
{d\rho \over d\zeta} - p_\perp \cos(\phi-\varphi) {d {\tilde \tau}
\over d\zeta} \right]
\rho(\zeta){\tilde \tau}(\zeta) d\zeta d\alpha_\parallel d\phi.
\label{sigmapbinv}
\end{equation}

The Cooper-Frye \cite{Cooper:1974mv} formalism yields the following expression for 
the momentum density of a given species of particles
\begin{eqnarray}
{dN \over dy d^2p_\perp} &=& {1 \over (2\pi)^3}
\int\limits_0^{2\pi} d\phi 
\int\limits_{-\infty}^{\infty} d\alpha_\parallel \int\limits_0^1 d\zeta
\,\, \rho(\zeta) {\tilde \tau}(\zeta) \nonumber \\
& \times & \left[m_\perp 
\hbox{cosh}(\alpha_\parallel-y) {d\rho \over d\zeta} 
- p_\perp \cos(\phi-\varphi) {d{\tilde \tau} \over
d\zeta} \right] \nonumber \\
& \times & \left\{
\exp \left[\beta m_\perp \hbox{cosh}(\alpha_\perp(\zeta)) 
\hbox{cosh}(\alpha_\parallel-y) - 
\right. \right. \nonumber \\
& & \left. \left. 
\beta p_\perp  
\hbox{sinh}(\alpha_\perp(\zeta)) \cos(\phi-\varphi) - \beta \mu \right] 
\pm 1 \right\}^{-1},
\label{dNdydpt}
\end{eqnarray}
where $\beta=1/T$ and $\mu=B \mu_B + S \mu_S +I_3 \mu_{I_3}$. The expression in curly
brackets is the Bose-Einstein or Fermi-Dirac distribution, with the corresponding
minus or plus signs.
{\tt THERMINATOR} implements two models of the freeze-out hypersurface and velocity field, 
described in the following subsections.

\subsection{The Cracow single-freeze-out model}
\label{sect:cracow}

We term the model \cite{Broniowski:2001we}, that has been used very successfully in the
description of the RHIC data, the Cracow single-freeze-out model.  
It is defined by the equations
\begin{equation}
{\tilde \tau} = \tau \, \hbox{cosh}(\alpha_\perp), \quad
\rho = \tau \, \hbox{sinh}(\alpha_\perp), \quad
\tau = \hbox{const}.
\label{CFM}
\end{equation}
and $u^\mu=x^\mu/\tau$. 
This parameterization yields 
the following distribution of the primordial particles 
\begin{eqnarray}
&&\!\!\!\!\!\!\!\!\!\! {dN \over dy d\varphi p_\perp dp_\perp d\alpha_\parallel 
d\phi \rho \, d\rho } = {1 \over (2\pi)^3}
\left[m_\perp \sqrt{\tau^2+\rho^2} \, \hbox{cosh}(\alpha_\parallel-y) 
- p_\perp \rho \cos(\phi-\varphi) 
\right] \nonumber \\
&& \!\!\!\!\!\!\!\! \times \left\{
\exp\left[\beta m_\perp  \, \sqrt{1+ {\rho^2 \over \tau^2}} 
\hbox{cosh}(\alpha_\parallel-y) - \beta p_\perp  
{\rho \over \tau} \cos(\phi-\varphi) - \beta \mu \right] 
\pm 1 \right\}^{-1}\!\!\!\!\! .\nonumber \\
\label{modA3}
\end{eqnarray}
The variable $\rho$ is limited from above by the maximum transverse size 
$\rho_{\rm max}$.

\subsection{Blast-Wave model}
\label{sect:blastwave}

The other considered model of the fire-ball expansion is the Blast-Wave 
model~\footnote{We should stress that the meaning of the blast-wave model 
here is different from the popular 
usage in fits to the heavy-ion data, since {\em we include the decays of resonances.}}.
In this case one specifies the freeze-out hypersurface and the velocity
profile of the flow by the conditions:
\begin{equation}
{\tilde \tau} = \tau = \hbox{const}, \quad
v_r = \hbox{tanh} \alpha_\perp (\zeta) = \hbox{const}.
\end{equation}
Then Eq. (\ref{dNdydpt}) takes the form
\begin{eqnarray}
&&{dN \over dy d \varphi p_\perp d p_\perp d\alpha_\parallel d\phi \rho d\rho} = 
{\tau \over (2\pi)^3} 
\, m_\perp \hbox{cosh}(\alpha_\parallel-y) \nonumber \\
& & \times \left\{
\exp\left[ {\beta m_\perp  \over \sqrt{1 - v_r^2}}
\hbox{cosh}(\alpha_\parallel-y) -
{ \beta p_\perp  v_r \over \sqrt{1 - v_r^2}} \cos(\phi-\varphi) - \beta \mu \right] 
\pm 1 \right\}^{-1}. \nonumber \\
\label{modA3bis}
\end{eqnarray}
Again, the variable $\rho$ is limited from above by the maximum transverse size 
$\rho_{\rm max}$.

\subsection{Decays of resonances}
\label{resdec}

We incorporate all the four-**** and three-*** resonances.
Following the scheme implemented in {\tt SHARE} \cite{Torrieri:2004zz} we
exclude all single-* resonances, and
practically all double-** resonances listed in the Particle Data Tables
\cite{Hagiwara:2002fs}. The reader may use his preferences by modifying the input files.
The stable particles and resonances are populated according to the 
formulas (\ref{modA3}) or (\ref{modA3bis}). All particles have 
their positions and momenta specified. Then, each resonance decays after some time, 
controled by its lifetime $1/\Gamma$. In the rest frame of the resonance the
decay at time $\tau$ occurs with the probability density $\Gamma \exp(-\Gamma \tau)$.
The decays are two-body or three-body, and the physical values of the 
branching ratios are taken from 
the Particle Data Tables. Heavy resonances may decay in cascades.
The kinematics of the decay is implemented in the rest frame of the 
decaying particle, and then the system is boosted to the 
reference frame in which the freeze-out hypersurface of Eqs.~(\ref{modA3})
and (\ref{modA3bis}) is defined.
For collisions of symmetric 
nuclei at RHIC it corresponds to the center-of-mass (CMS) frame.  

The particles are distinguished by isospin, hence the 
branching ratios for strong decays incorporate the isospin Clebsch-Gordan factors.
The method of computing the branching ratios, the policy of assigning the 
branching ratios to channels labeled as {\em seen} or {\em dominant}, and 
other details concerning the decays of particles are incorporated in the same 
fashion as in {\tt SHARE} \cite{Torrieri:2004zz} where the reader is referred to for 
more details.

\section{Description of THERMINATOR}

{\tt THERMINATOR} is written in {\tt c++} and uses {\tt ROOT} -- {\em An object-oriented 
data analysis framework} \cite{root}. A more sophisticated usage of {\tt THERMINATOR} at the level of 
analysing the output requires a basic knowledge of {\tt ROOT} from the user.  

\begin{figure}[h]
\begin{center}
\includegraphics[width=12.5cm]{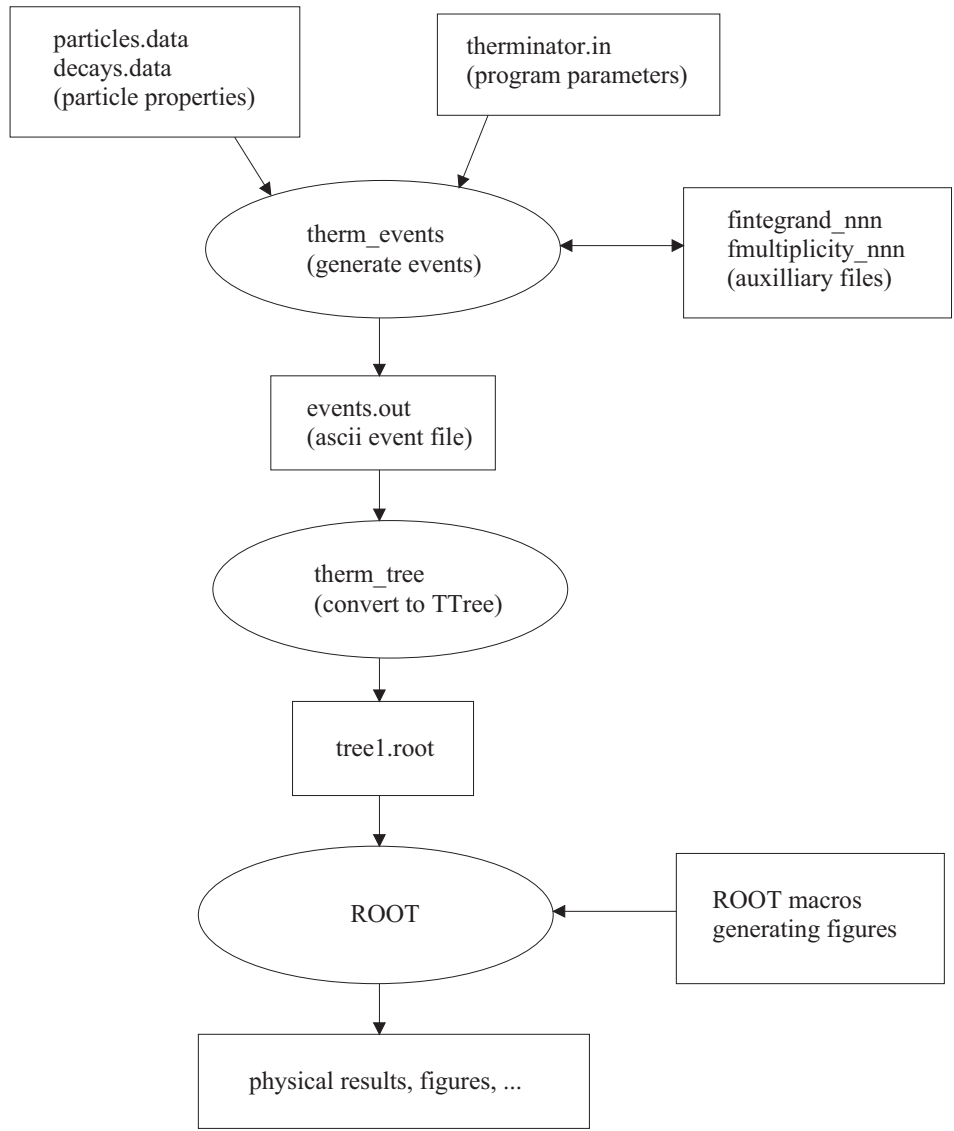}
\end{center}
\caption{The block structure of {\tt THERMINATOR}.}
\label{block}
\end{figure}

The block structure of {\tt THERMINATOR} is shown in Fig.~\ref{block}.
Information on particle properties as well as the {\tt THERMINATOR} parameters are read from the 
(modifiable) input files. Two programs, {\tt term\_events} and {\tt term\_tree} 
generate a {\tt ROOT} tree which can be used within {\tt ROOT} to generate 
physical results. The blocks of Fig.~\ref{block} are described in detail in the following.

\subsection{Particle data files ({\tt particles.data}  and  {\tt decays.data})}

The structure of input files containing the information on the particle
properties and their decays is the same as in {\tt SHARE} \cite{Torrieri:2004zz}, where the
reader is referred to for a more detailed description. Here we just give the 
basic information.

The input file {\tt particles.data} contains the information on the properties of 
particles such as mass, width, spin, isospin, the quark contents, and the
Monte Carlo identification number.  The file has the
following format:

\vspace{0.1cm}

\noindent {\bf name\quad mass\quad  width\quad  spin\quad  I\quad  
I3\quad  q\quad  s\quad  aq\quad  as\quad  c\quad   ac\quad  MC}

\vspace{0.1cm} 

\noindent where:

\begin{description}
\setlength{\itemsep}{-0.1cm}
\setlength{\labelwidth}{1.5cm}
\setlength{\itemindent}{0.5cm}
\item[name] the particle label used in the program,
\item[mass] mass in GeV,
\item[width] width in GeV,
\item[spin] spin,
\item[I] isospin,
\item[I3] 3rd component of isospin,
\item[q,\ s\hfill] number of light/strange quarks,
\item[aq,\ as\hfill] number of light/strange antiquarks,
\item[c,\ ac\hfill] number of charm/anticharm quarks,
\item[MC] particle's identification number, where
  available corresponding to the standard Monte Carlo particle
  identification convention \cite{Hagiwara:2002fs}.
\end{description}

\bigskip

The file {\tt decays.data} contains  the information on particle decays in the format:

\vspace{0.3cm}

\noindent {\bf Name$_{\rm parent}$  Name$_{\rm daughter1}$  Name$_{\rm
    daughter2}$  Name$_{\rm daughter3}$\footnote{
Appears for the three-body decays only.}  BR
    C--G?(0/1)}

\vspace{0.3cm}

\noindent where {\bf BR} denotes the branching ratio of the decay
and {\bf C--G} refers to whether the branching ratio should be multiplied by a
Clebsch--Gordan coefficients ({\bf 0}:~no, {\bf 1}: yes). Normally, we use {\bf 1} for the two-body 
decays and {\bf 0} for the three-body decays, where the full (isospin-dependent) 
branching ratio BR is provided. The entry BR may be used to control the feed-down from 
the weak decays, in particular setting it to 0 would switch off the decay channel.

\subsection{Input parameter file ({\tt therminator.in})}
\label{inpfile}

The execution of the {\tt THERMINATOR} program is mainly controled by the
parameter input file, whose default name is {\tt therminator.in}. Different parameter 
filename can be specified by the user at the command line as a parameter to the program.
The file contains the following information, which must be filled by the user:
\begin{description}
\setlength{\itemsep}{-0.1cm}
\setlength{\labelwidth}{1.5cm}
\setlength{\itemindent}{0.5cm}
\item[NumberOfEvents] -- the number of events to be generated (default: 500), 
\item[Randomize] -- if set to 1, the random number generator seed is
  initialized with the help of the current time, producing different
  seeds for every run of the program; if set to 0 the seeds (and
  therefore all the results) will be the same for each run of the program (default: 1),
\item[InputDirSHARE] -- location of the {\tt SHARE} input files (default: ../share),
\item[EventOutputFile] -- name of the ASCII file containing the generated events (default: event.out),
\item[FreezeOutModel] -- selects the version of the freeze-out model used, the ``Cracow'' 
model (see Sect. \ref{sect:cracow}) (value: {\tt
SingleFreezeOut}) as well as  ``BlastWave'' model 
   (see Sect. \ref{sect:blastwave}) (value: {\tt BlastWaveVT}) are
available (default: {\tt SingleFreezeOut}) ,
\item[BWVt] -- a parameter specific to the Blast-Wave model - the
common velocity (default: 0.55) ($v_r$ from (\ref{modA3bis})),
\item[Tau, RhoMax] -- geometric parameters in units of fm
(default: 9.74, 7.74) ($\tau$ from (\ref{modA3}) or (\ref{modA3bis})
and $\rho_{max}$ from Sec. \ref{sect:cracow} or \ref{sect:blastwave}),

\item [Temperature, MiuI, MiuS, MiuB] -- thermodynamic parameters in units of GeV
(default: 0.1656, -0.0009, 0.0069, 0.0285) ($T$ from (\ref{modA3}) or
(\ref{modA3bis}) and $\mu_{I_{3}}$, $\mu_B$, and $\mu_S$ from Sec. \ref{sect:basics}),

\item[AlphaRange, RapidityRange] -- integration ranges for
  longitudinal variables (space-time rapidity and rapidity) (default:
8.0 (-4.0, 4.0), 4.0 (-2.0,2.0) ) ($\alpha_{\parallel}$ and $y$ from (\ref{modA3}) or (\ref{modA3bis})),

\item[NumberOfIntegrateSamples] -- number of samples used in the determination of multiplicity
and maximum of the integrand (default: 1000000).
\end{description}

The parameter file is properly commented for the ease of use.

\subsection{Generation of the particle distribution}

The generation of particle distributions proceeds in  three main
steps, two of which are performed only once per given 
parameter set.

In order to generate particles through a Monte Carlo method, the maximum
value of the distributions on the right-hand-side of Eq. (\ref{modA3}) or 
(\ref{modA3bis}) must be known. It is found through a simple numerical procedure. 
A sample of particles is generated and the values of the distributions
are
calculated for each of them. The maximum value obtained is taken as the maximum
of the requested distribution for the considered particle type. The authors have checked, by 
studying the resulting distributions, that if the size of
the sample is large enough, the method provides a good estimate of the maximum
value. The maximum value depends, in principle, on the particle type 
and values of parameters, but does not change from
event to event. Therefore the value is calculated once for each
particle type and stored in the file {\tt fintegrandmax\_nnn}, where {\tt nnn}
stands for a unique identifier for a given parameter set. Subsequent
generations of events with the same parameters do not require the
regeneration of the maximum values; they are read from the {\tt fintegrandmax\_nnn} file
instead. This way the computation time is saved.

In order to generate events, a multiplicity of each particle type must
be known. The average multiplicity per event can be calculated in a
straightforward manner by numerically integrating the distribution
functions in the given integration ranges (determined by the model
parameters). This procedure must also be done only once per parameter
set. The average multiplicity of each particle is stored in the
working directory in a file named {\tt fmultiplicity\_nnn}, where {\tt nnn}
is the same unique identifier, as the one used for the 
maxima discussed above. The multiplicities are then read in by subsequent
generations, saving the time of calculation.

In the next step the program proceeds to its ultimate goal of generating events, 
{\em i.e.}, the data sets containing
full information about produced particles, the history of their evolution, resonance decays, 
{\em etc.} Each event is generated separately. First, 
the multiplicities of each particle type are generated as random numbers from a Poissonian 
distribution, with the mean being the average particle multiplicity determined earlier. 
Then the program proceeds to generate particles, sequentially  from the heaviest to the 
lightest particle type. The procedure is, in essence, the generation of the set of six 
random numbers (the magnitude of the transverse momentum $p_{T}$, 
the azimuthal angle of the transverse momentum $\varphi$, the rapidity $y$ associated with the particle's
momentum, the space-time
rapidity $\alpha_\parallel$ associated with the particle's location, 
the transverse distance of the particle $\rho$, the azimuthal angle of the particle's location 
$\phi$), distributed 
according to the formulae ({\ref{modA3}}) or ({\ref{modA3bis}}), depending on the 
selected source model. The integration ranges are determined by the model parameters 
and the user's input. The event generation procedure is a standard von Neumann method of 
rejection/acceptance of the given set of numbers based on the randomly generated test 
value distributed uniformly between 0 and $f_{\rm max}$, where $f_{\rm max}$ is the 
maximum value of the distribution determined earlier. The accepted sets of numbers are 
stored in memory as representing the actual particles. The procedure ends when the 
determined number of particles of each particle type is
generated. At this point all the {\em primordial} particles, stable and resonances, 
have been generated and stored in the event.

\subsection{Implementation of two-body decays}

The next step in the event generation is the simulation of the decays of unstable
particles. A particle is treated as unstable if it has non-zero width
in the {\tt particles.data} input file. If the user wishes to prevent a decay
of a particular resonance, he should
modify the input file and set the width of a given particle to 0. Alternatively,
decay channels may be switched-off selectively by setting the corresponding branching 
ratio in {\tt decays.data} to zero. 

The decays proceed sequentially from the heaviest to the lightest
particle. The decay daughters are immediately added to the set of
particles in the current event, hence they may decay in the 
subsequent steps. 
Most particles have several decay channels.
In each decay one of them is selected randomly with appropriate probability
corresponding to the branching ratio from the
{\tt decays.data} input file. It may happen that the selected decay
channel is of a sub-threshold type (average mass of the parent
particle is smaller than the sum of masses of the daughter particles),
which is not yet implemented in
{\tt THERMINATOR}. In such a case, the particle does not decay and the
program proceeds to the next particle.

The selected decay channel can be two- or three-body. Both are
implemented in {\tt THERMINATOR} and treated on equal footing. We turn now
to the discussion of the two-body decays;  the
specifics of the three-body decays are described in the next
subsection. 

First the lifetime $\Delta \tau$ of the decaying particle
of mass $M$, mowing with the four-momentum $p^\mu$, is generated 
randomly according to the exponential decay law, $\exp(-\Gamma \Delta \tau)$.
Once the lifetime $\Delta \tau$ 
of the particle is known, the space-time point of its decay can be calculated, by taking 
its original space-time position $x^\mu_{\rm origin}$ (determined from the values 
$\tau$, $\rho$, $\alpha_\parallel$, and $\phi$), and adding to it the space-time distance 
traveled by the particle,
\begin{equation}
x^\mu_{\rm decay} = x^\mu_{\rm origin} + {p^\mu \over M} \Delta \tau.
\label{decpt}
\end{equation}
This calculation is always done in the CMS reference frame of the colliding system
in the Cartesian coordinates. Note that Eq. (\ref{decpt}) guarantees that the 
space-time interval between the origin and decay point is equal to $\Delta \tau$.

The energies $E_1$, $E_2$, and the three-momenta ${\bf p}_1$, ${\bf p}_2$ of the daughter 
particles are initially determined in the rest frame of the parent particle 
from the energy-momentum conservation laws, which yield: 
\begin{equation}
E_{1,2} = {{M^2 - m_{2,1}^2 + m_{1,2}^2} \over {2M}},
\label{decay1}
\end{equation}
\begin{equation}
|{\bf p}_1| = |{\bf p}_2| = 
{ [(M^2 -(m_1+m_2)^2) (M^2 - (m_1-m_2)^2)]^{1/2} \over {2M} }.
\label{decay2}
\end{equation}
Here $m_1$, $m_2$ are the masses
of the daughter particles. The direction of the three-momentum of particle $1$ is 
generated randomly and is evenly distributed on the sphere in the parent particle's rest
frame. The direction of the three-momentum of particle $2$ follows from the momentum
conservation. Having generated the decay momenta of both particles in the
resonance rest frame, we next boost them to the CMS frame with the parent particle
velocity. The  decay point of the parent particle (\ref{decpt}) is taken as the
origin of the daughter particles.

Together with each daughter particle the reference to the parent
particle is stored. This enables the full tracing of the cascade
decay tree in the analysis of the {\tt THERMINATOR} output. Also, the
parent particle is flagged as ``decayed'' which is later indicated in
the output file.

\subsection{Implementation of three-body decays}

Following the common practice, we assume that the transition amplitude in three-body
decays is constant, {\em i.e.}, not dependent on the momenta. Similarly to the case of two-body
decays, the momenta of the emitted particles are first determined in the parent particle's 
rest-frame and only at the end are boosted to the CMS 
frame with the parent velocity. The three-momenta of daughter particles considered in the 
parent particle rest frame lie all in one plane. Moreover, the energy distributions of the 
daughter particles in three-body decays with constant matrix element are uniform
(see for example Ref. \cite{Weinberg:1995mt}), therefore 
we may generate randomly the energies $E_2$ and $E_3$ of the daughter 
particles $2$ and $3$ in the energy range allowed by the energy conservation:
$m_2 \leq E_2 \leq M-m_1-m_3, m_3 \leq E_3 \leq M-m_1-m_2$. 
The energy of the daughter particle $1$  as well as the magnitudes of all three-momenta 
($|{\bf p}_1|$, $|{\bf p}_2|$, and $|{\bf p}_3|$)  follow from the energy and
momentum  conservation laws, which in addition determine the angle between the 
vectors ${\bf p}_2$ and ${\bf p}_3$,
\begin{eqnarray}
E_1 &=& M - E_2 - E_3, \nonumber \\
M &=& \sqrt{ {\bf p}_2^2 + 2 |{\bf p}_2| |{\bf p}_2| \cos\theta_{23} +  
{\bf p}_3^2 + m_1^2} + \sqrt{|{\bf p}_2|^2 + m_2^2} + \sqrt{|{\bf p}_3|^2 + m_3^3 }.
\nonumber \\
\label{3d}
\end{eqnarray}
In our approach Eq. (\ref{3d}) is used to calculate $\cos\theta_{23}$ and whenever
the condition $|\cos\theta_{23}| \leq 1$  is satisfied the generated values of
the energies and the angle itself are accepted. In this case,  we rotate the
plane spanned by the vectors ${\bf p}_2$ and ${\bf p}_3$, together with the whole system, 
by a randomly chosen angle $\phi_{23}$ around the axis shown by the vector ${\bf p}_3$. 
Next, with the help of two additional randomly
generated Euler angles the orientation of the vector ${\bf p}_3$ is chosen. Finally,
the system is boosted with the velocity of the parent particle.
Obviously, the generated values of  $|{\bf p}_1|$, $|{\bf p}_2|$, and $|{\bf p}_3|$ 
for which no solution of  Eq. ({\ref{3d}) exists are rejected.

Similarly to the two-body decays,  the space-time origin of daughter
particles is taken as the decay point of the parent. The daughters are
automatically added to the list of particles in the event.

\subsection{Storage of events}
\label{sect:stor}

The particles are stored in the computer memory as collections
corresponding to one event. Once the whole event is generated (all
primordial particles are produced and all the unstable particles have
sequentially decayed), the event is written out to the output file and
is erased from the memory.

Each event in the output file consists of an event header (containing
the number of particles in the event), and the particles
themselves. Each particle is represented in a single line. It
contains: the particle index for this event (useful for
cross-referencing the decay parents and daughters), Particle Data Group identification
number, PID, components of the momentum:
$p_x$, $p_y$, $p_z$, energy $E$, mass, $m$ (in GeV), space-time coordinates of creation: 
$x$, $y$, $z$, $t$ (in
1/GeV), the parent index (-1 if the particle is primordial), finally, the decay flag
(1 if particle has decayed, 0 if not). An example of the output file is shown in Fig.
\ref{outfex}.

The format of the output file is a plain ASCII text, which is a standard used
by many event generators. It has the advantage of having the complete
information and being easily readable by the human or a computer program. It is,
however, not efficient in terms of disk-space usage and data-access
time. Therefore a simple {\tt ROOT} macro {\tt therm\_tree.C} is
provided, which converts the text output file into a {\tt ROOT TTree}, which
is stored in a {\tt ROOT} file on disk. Events are divided into 500-event
batches, each batch in a separate file. 

The information stored in {\tt ROOT} file differs slightly from
the one stored in the text file; it does not contain the index. The
position coordinates $x$, $y$, $z$, and $t$, are now converted to fm (we use $c$=1).
Instead of the parent index and decay flag the {\tt TTree} contains two PIDs. 
The first one, ``fatherpid'', is the Particle Data Group PID of the immediate parent particle
(or the PID of the particle itself, if it is primordial). The second,
``rootpid'', is the PID of the original primordial particle,
that is the particle at the beginning of the cascade decay
tree. This enables us to distinguish between three scenarios: 
1)~the primordial particle has ``pid'' = ``fatherpid'' = ``rootpid'',
2)~the particle coming from the decay of primordial particle has
``rootpid'' = ``fatherpid'' and its own, different ``pid'', and finally
3)~the particle coming from a cascade decay has its own
``pid'', a PID of its immediate parent -- the ``fatherpid'', and the PID of
the original primordial resonance -- ``rootpid''.


\section{Programming structure of {\tt THERMINATOR}}

\subsection{Technology and general design}

{\tt THERMINATOR} has been designed as an object-oriented program and has
been written in the {\tt c++} language. It also uses {\tt ROOT} classes to
store and manipulate its data structures. 

The classes can be divided into three functional blocks. The first one
deals with reading input files and storing the information in memory
for the use by other classes. The second block takes care of storing the
information of the generated particles and events. Finally, the third block contains
the main logic of the program, as well as most of the mathematical formulas. It
deals with the overall program execution control (including parsing the
parameters), multiplicity calculation, particle generation, the decays,
as well as storing the results.

\subsection{Data storage block}

The data storage functional block consists of several {\tt c++} classes. The
topmost is the ParticleDB, which stores pointers to information about
all ParticleTypes and allows for easy searches by name or by
number. It is filled with the data from the input files by the Parser
class at the beginning of the {\tt therm\_events} program. 

The basic object for storing the particle type data is the
ParticleType class. It contains all the characteristics which are
common to all the particles of the same type: the charge, baryon
number, mass, width, name, {\em etc.} It also contains the pointer to the
complete decay table of this particular particle type. There are as
many instances of the ParticleType class as there are valid particle
types read from the input file {\tt particles.data}.

The DecayTable class is a collection of pointers to the specific decay
channels. It deals consistently with the branching ratios. Its job is
also to select a given decay channel, based on the random number that
is passed to it.

The DecayChannel is a simple storage class, which contains the
identifiers of the daughter particle types as well as the values of the
branching ratios for particular decay channels.

\subsection{Calculation results storage block}

This block of classes deals with storing the results generated by the {\tt therm\_events}
program. It consists of two classes. The general one is the Event
class. It stores pointers to the particles contained in the event. It
is responsible for storing them sorted according to mass. It also
manages the event generation process and holds the generated particle
multiplicities for this specific event. 

The Particle class stores information on each generated
particle. There are as many instances of this class at any given
moment, as there are particles in a current event. Each particle is
defined by its particle type (which is a pointer to the corresponding
instance of the ParticleType class) and characteristics that vary from
particle to particle, that is the momentum ${\bf p}$,
the creation point ${\bf x}$ as well as information if it comes
from a decay and if so, from which particle, and if it has decayed
already. This class is also responsible for formatting the lines in the
output file.

\subsection{Main program block}

This block contains the main acting classes of the {\tt therm\_events} program. One of them
is the main program, which takes care of the command-line
parsing, instantiating the ParticleDB classes, initiating reading of
parameters and inputs, and running the appropriate generation and
calculation routines.

The Integrator class contains the most important part of physics
implementation of {\tt THERMINATOR}. It includes the procedure which
implements the Monte Carlo integrator of the emission functions. The
integrator has three modes of operation: search for the maxima of 
distributions, straightforward integration (for multiplicity
calculation), and generation of random numbers according to the
probability density described by a given emission function. This
latter procedure is the particle generation of {\tt THERMINATOR}. All of
these modes of operation are properly abstract and are not dependent
on the particular emission function form. Thanks to this, the
emission functions themselves can be relatively easily
modified and new ones can be added with little effort.

One more helpful class included in the main program block is the
ParticleDecayer. Its sole purpose is the implementation of the decay
process. It assigns the pointer to the father particle. It is
responsible for selecting (by the Monte Carlo method) the decay channel,
performing the necessary kinematic and space-time calculation (again
using the Monte Carlo procedures wherever it is appropriate, {\em e.g.}, in generating
randomly the directed decay axis). Finally it returns pointers to the 
newly created Particle class instances containing the daughter
particles. Both 2-body and 3-body decays are implemented in 
ParticleDecayer. 

\begin{figure}[h]
\begin{center}
\includegraphics[width=\textwidth]{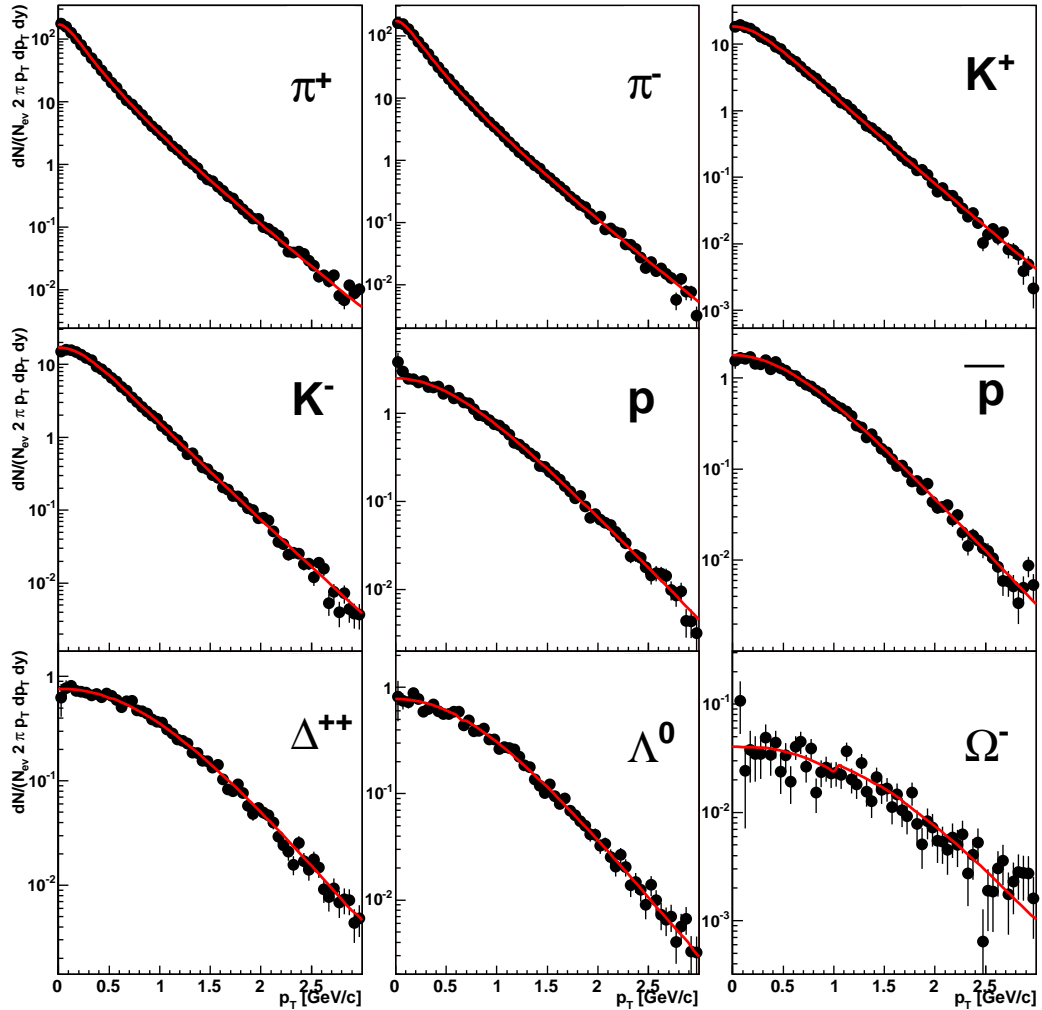}
\end{center}
\caption{Comparison of the $p_{T}$ spectra of primordial particles obtained
from {\tt THERMINATOR} (circles) and {\tt MathSHARE} \cite{mathsh} (lines)
for the Cracow single-freeze-out model. The model
parameters corresponding to the best fit to the most central PHENIX data
for Au + Au @ 200 GeV are taken from Ref. \cite{Baran:2003nm}.
The sample contains 1000 events. The errors on the {\tt THERMINATOR} results
are statistical. A perfect agreement of the two independent codes is observed. }
\label{thshcmpnd}
\end{figure}

\begin{figure}[tb]
\begin{center}
\includegraphics[width=\textwidth]{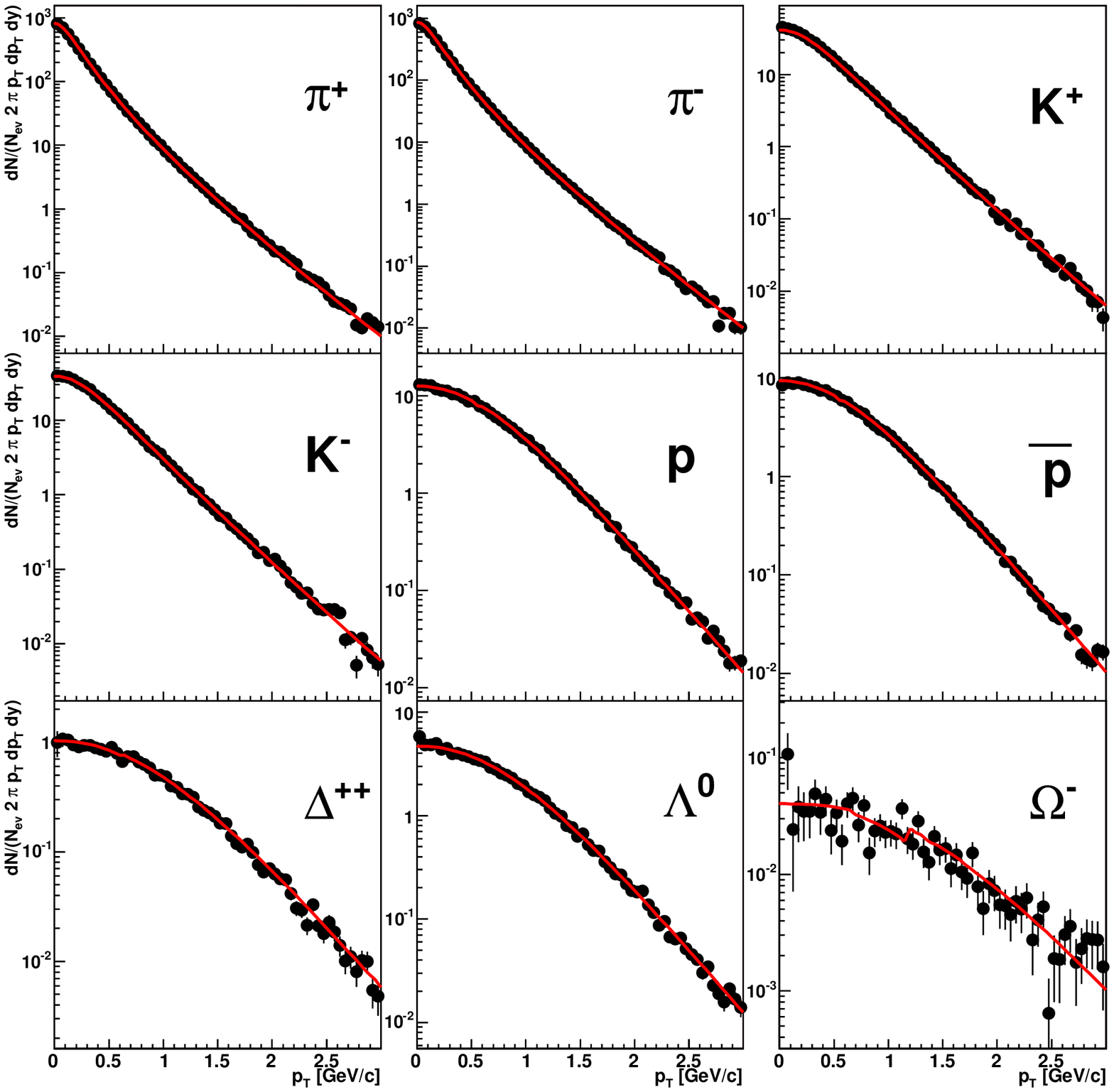}
\end{center}
\caption{Same as Fig. \ref{thshcmpnd} for the final (primordial
+ products of decays) distributions. A perfect agreement of the two independent codes 
is visible. }
\label{thshcmpdc}
\end{figure}

\section{Testing against existing calculations}

During the development of {\tt THERMINATOR}, special care was taken to
ensure that the obtained results are in perfect agreement with those
delivered by the {\tt MathSHARE} package \cite{mathsh}. The most sensitive test
available was to check the shape of the $p_{T}$-spectra of many 
different particle species. Since {\tt THERMINATOR} is based on
Monte Carlo methods, the agreement could only be judged within the
uncertainties of such method, which are decreasing with the increase in
the number of generated events. 

The agreement at the input level was ensured by simply using exactly 
the same files {\tt particle.data} and {\tt decays.data}
as input to {\tt THERMINATOR} and {\tt MathSHARE}. In this way,
the same number and types of particles were used, as well
as the same decay channels and branching ratios. The tests were done in
two steps. First, the $p_{T}$-distributions of primordial particles were
compared. The results of this comparison are seen in Fig.~\ref{thshcmpnd}. For all 
considered particle types, the results of quasi-analytic calculations
done in {\tt MathSHARE} and histograms obtained with the help of 
{\tt THERMINATOR}
agree within the error expected from the Monte Carlo procedure. The second 
test was the analysis of the $p_{T}$-spectra, but this time with full resonance 
contribution included. The results for this test can be seen in Fig.
{\ref{thshcmpdc}}. Again, the agreement reached with a sample of 
1000 events is impressive, proving that
not only the initialization procedures but also the decay procedures 
yield identical results.

\section{Yielding new results\label{sec:new}}

\begin{figure}[b]
\begin{center}
\includegraphics[width=9cm]{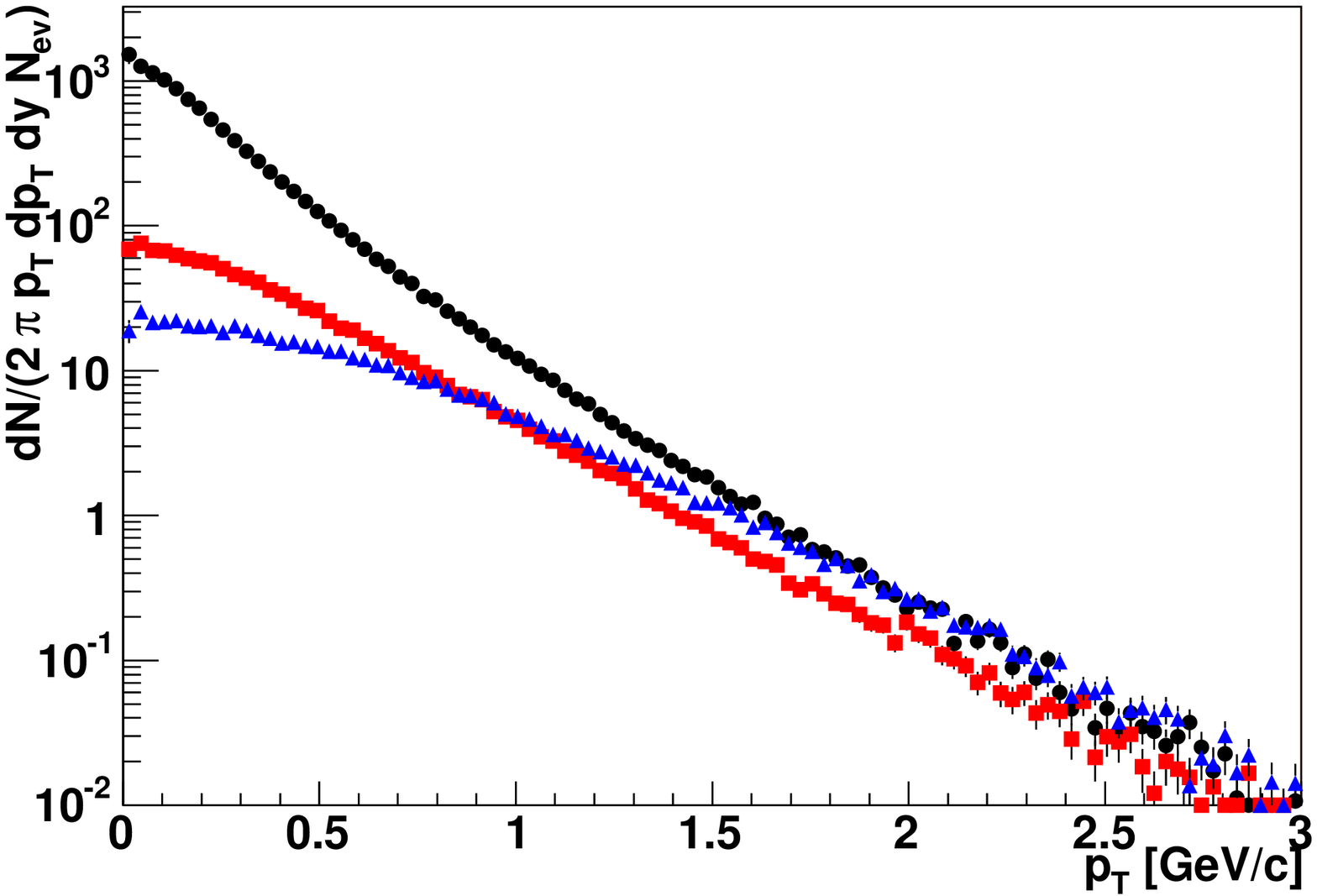}
\end{center}
\caption{The $p_{T}$ spectra of pions (circles), kaons (squares) and 
protons (triangles) generated from {\tt THERMINATOR} for the Cracow single-freeze-out model
with default parameters from {\tt therminator.in}.
The sample contains 500 events. The errors are statistical. The figure is generated by the {\tt ROOT}
script {\tt figure5.C}.}
\label{ptsp}
\end{figure}

This Section contains examples of the physical results obtained with the help of {\tt THERMINATOR}. 
More results and details focusing on physics will be presented in future publications. 
All the simulations were done with the parameters taken from Ref.~\cite{Baran:2003nm}, 
which were obtained by fitting the STAR experiment data with the {\tt MathSHARE}
package. These parameters are included as default in the input file
{\tt therminator.in}. The generated sample used here contains 500 events.

\begin{figure}[t]
\begin{center}
\includegraphics[width=9cm]{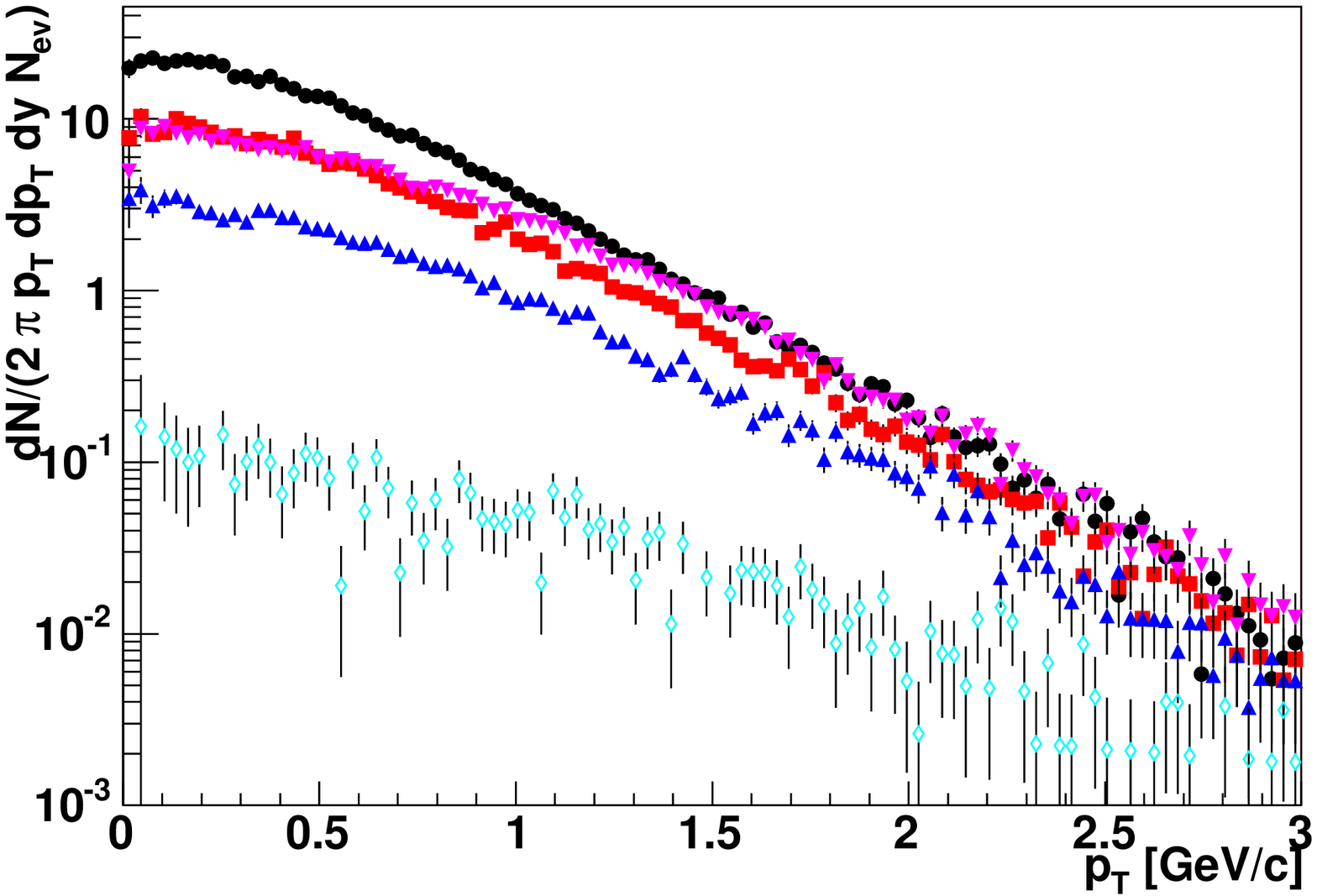}
\end{center}
\caption{The $p_{T}$ spectra of $\rho$ mesons (filled circles), $K^*_0$
mesons (squares), $\phi$ mesons (up-triangles), $\Lambda^0$ barions (down-triangles),
and $\Omega^-$ barions (diamonds) generated from {\tt THERMINATOR} for the Cracow single-freeze-out model
with default parameters from {\tt therminator.in}.
The sample contains 500 events. The errors are statistical. The figure is generated by the {\tt ROOT}
script {\tt figure6.C}.}
\label{ptsp1}
\end{figure}

In Figs. {\ref{ptsp}}  and {\ref{ptsp1}} we show the $p_{T}$-spectra of various particle species.
Such spectra were first reported in the original papers 
on the single-freeze-out model \cite{Broniowski:2001we,Broniowski:2001uk}.
The user may generate his own spectra for any desired particle 
by means of simply modifying the attached ROOT scripts {\tt figure5.C} or {\tt figure6.C}. 

Figure \ref{yspec} shows the rapidity distributions of pions, kaons, and protons in the
midrapidity region. One may observe the boost-invariance imposed on the model.
With relatively simple modifications the program may be used to describe realistic
rapidity distribution in the full rapidity range, such as for example in Ref.
\cite{Florkowski:2004tn}.

{\tt THERMINATOR} allows for deeper understanding of the physical results, in particular
of the role played by the resonances. As an example in Fig. \ref{pires} we present the
anatomy of the $\pi^+$ transverse momentum spectra. One can see that the contribution 
from the resonances is dominant, with only one quarter of all pions being primordial.
The resonance effect is important in the whole displayed $p_T$ range. One can also
see that, as expected from the kinematics, the contribution from the three-body decay
of the omega meson has a steeper distribution than the contributions from two-body
decays. Analogous anatomy of the proton $p_T$ spectra is presented in Fig. \ref{pires1}.

\begin{figure}[tb]
\begin{center}
\includegraphics[width=9cm]{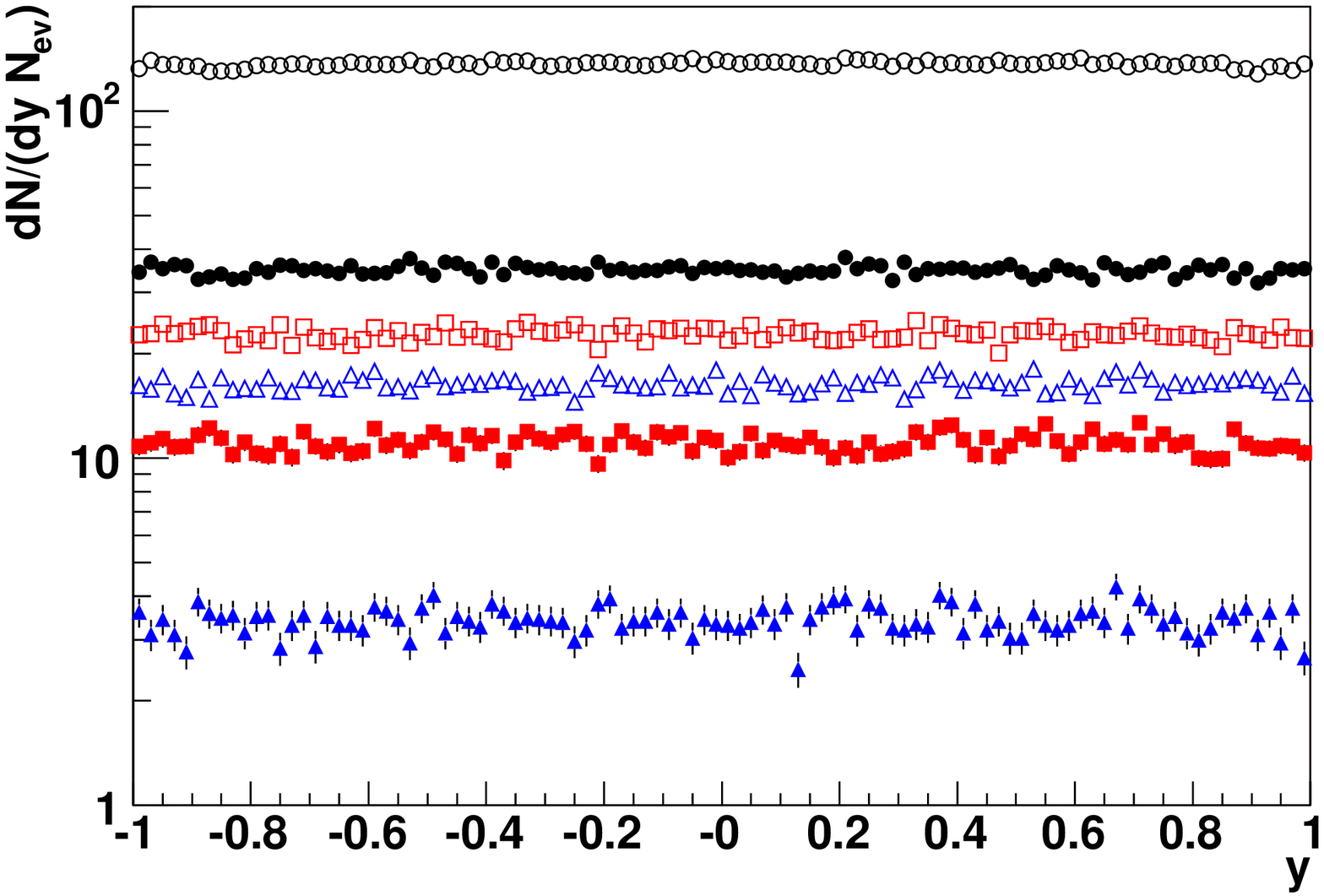}
\end{center}
\caption{The rapidity distributions of pions (circles), kaons (squares) and protons (triangles) 
generated from {\tt THERMINATOR} with the default input file. The
closed symbols show the rapidity distribution of primordial particles.
The open symbols show the distributions of the final particles. 
The sample contains 500 events. The boost-invariance assumed in the present model
is clearly seen. The figure is generated by the {\tt ROOT}
script {\tt figure7.C}.}
\label{yspec}
\end{figure}

\begin{figure}[tb]
\begin{center}
\includegraphics[width=9cm]{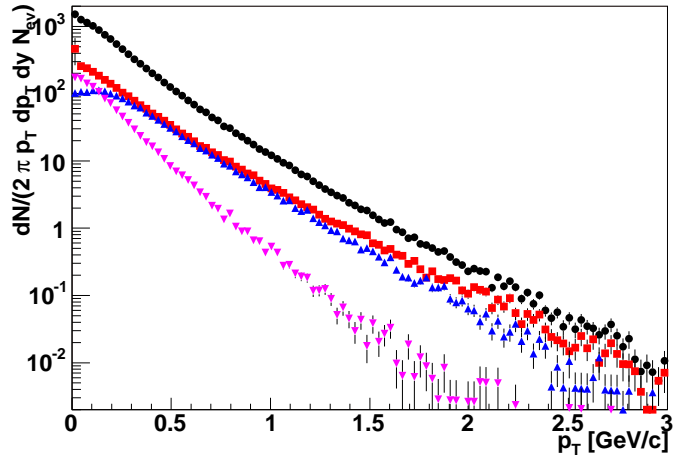}
\end{center}
\caption{The anatomy of resonance contribution to the $\pi^+$ transverse-momentum
spectra. The plot shows absolute contribution to the overall $\pi^+$
spectra (circles) from: primordial pions (squares), pions from the
$\rho^0$ and $\rho^+$ decays (up-triangles), and pions from the $\omega$
decays (down-triangles). The sample contains 500 events. The figure is generated by the {\tt ROOT}
script {\tt figure8.C}.}
\label{pires}
\end{figure}

\begin{figure}[tb]
\begin{center}
\includegraphics[width=9cm]{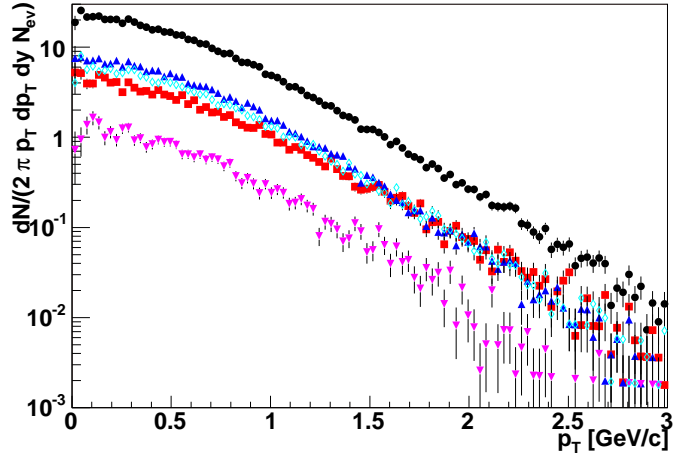}
\end{center}
\caption{The anatomy of resonance contribution to the proton $p_T$
spectra. The plot shows absolute contribution to the overall proton
spectra (closed circles) from: primordial protons (squares), protons from the
$\Lambda^0$ decays (up-triangles), protons from the $\Sigma$ hyperon
decays (down-triangles), and protons from the $\Delta$ resonance
decays (open circles). The sample contains 500 events. The figure is generated by the {\tt ROOT}
script {\tt figure9.C}.}
\label{pires1}
\end{figure}

\begin{figure}[tb]
\begin{center}
\includegraphics[width=0.9\textwidth]{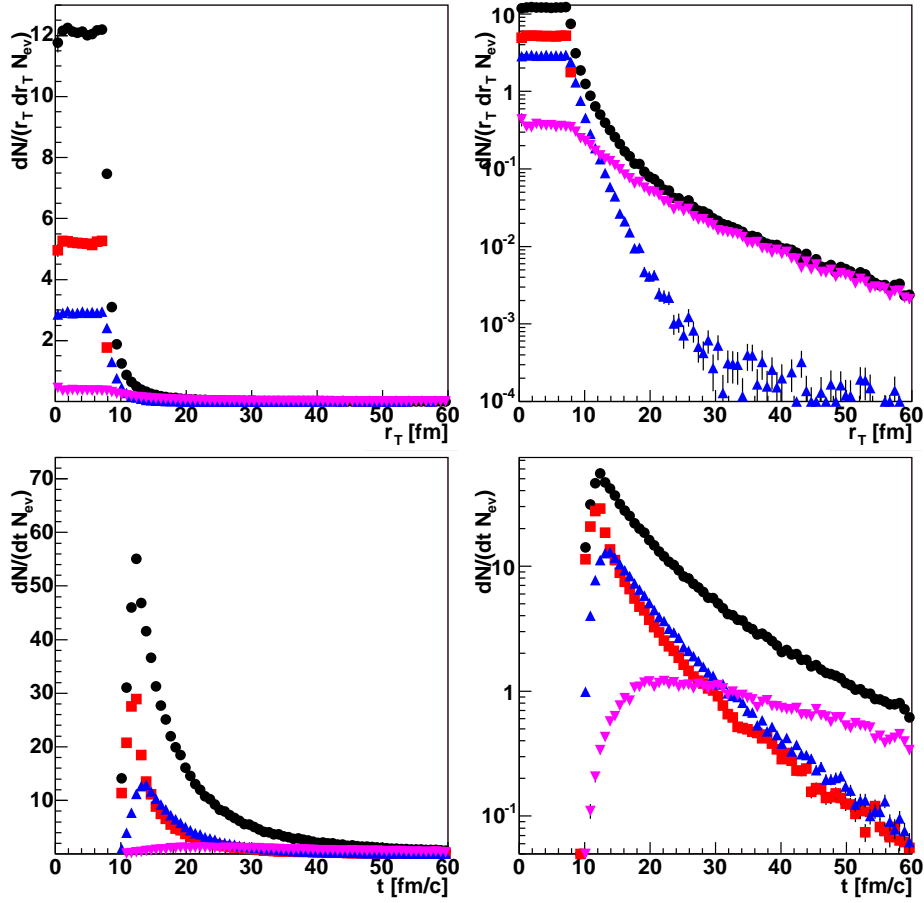}
\end{center}
\caption{Space-time picture of the $\pi^+$ emission process. The plots
in the upper row show spatial distribution of the pion emission
points, $dN/(r_T dr_T)$.
The symbol $r_T$ denotes the transverse radial
coordinate of the points where the pions are born. The lower plots
show the distribution of the pion emission times. Plots on the left are with
linear vertical scale, plots on the right are copies with the
logarithmic scale. The plots show absolute
contribution to the overall distributions 
(circles) from the primordial pions (squares), the pions from the
$\rho^0$ and $\rho^+$ decays (up-triangles), and the pions from the $\omega$
decays (down-triangles). The sample contains 500 events. The figure is
generated by the {\tt ROOT} script {\tt figure10.eps}, see the text for more details.}
\label{srad}
\end{figure}

\begin{figure}[tb]
\begin{center}
\includegraphics[width=9cm]{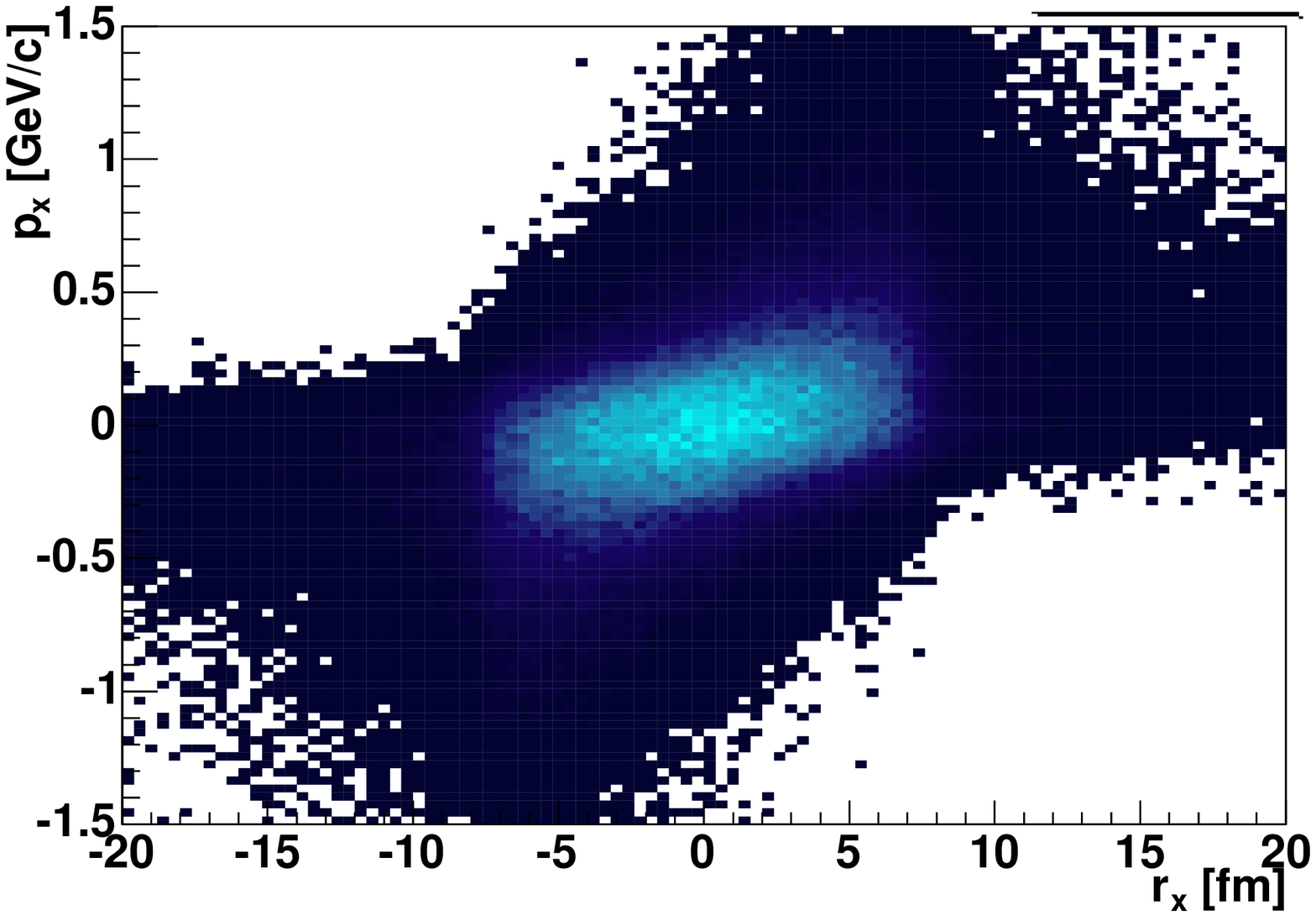}
\end{center}
\caption{The space-momentum correlation in THERMINATOR. The plot
shows the correlation between spatial and momentum x-coordinate for the
pions. The sample contains 500 events. The figure is generated by the
{\tt ROOT} script {\tt figure11.eps}}
\label{xpx}
\end{figure}

Further plots show some very interesting and novel observables
which were not accessible from previous programs implementing the model.
These observables refer to the space-time picture of particle production
in relativistic heavy-ion collisions.  For instance, 
in Fig. \ref{srad} we show the distribution of the space and time coordinates  
of the emission points of pions. For the primordial particles the emission point
is the point on the freeze-out hypersurface, while for particles coming from
resonances the emission point is simply the point where the decay occurs.
The top-left panel shows the distribution of
the transverse coordinates of the pion emission points, $r_T$.
The primordial pions are denoted by squares, the pions from the $\rho$ meson
by up-triangles, and the pions from the $\omega$ meson by down-triangles.
The dots indicate the full yield (from {\it all} resonances). 
For $r_T$ in the range from zero to $\rho_{\rm max} = 7.74$ fm
(the maximum transverse size of the freeze-out hypersurface) the observed
distributions are flat, reflecting the uniform distributions of the primordial
particles. For larger values of $r_T$ we can see the tails generated by the decays
of the resonances. Resonances with longer lifetime generate longer tails.
This is clearly seen in the top-right panel, where the logarithmic vertical scale
is used. One can also observe that the contribution from the long living $\omega$ meson
saturates the full contribution at $r_T > 20$ fm. 

The two lower panels of Fig. \ref{srad} show the analogous analysis with regard to
the emission time  $t$.  The emission starts at $t = \tau = 9.74$ fm and peaks
around $t = 14$ fm. This reflects the geometry of the freeze-out hypersurface
assumed in the Cracow model, 
Eq. (\ref{CFM}), as well
as the finite lifetime of the resonances. 

{\tt THERMINATOR} may be used to analyze various correlations.
As an example, Fig. {\ref{xpx}} shows the space-momentum correlation
encountered in relativistic heavy-ion collisions. One can see that
the $x$-coordinate of the pion emission point is correlated with the
$x$-coordinate of its momentum. This is a characteristic feature of the
systems exhibiting the transverse flow.

The plots shown in this paper are only a few examples of the variety of physical
results that can be obtained with {\tt THERMINATOR}. 
Many novel analyses may be
performed, including a detailed study of the resonance decay contribution to
particle spectra,  a comprehensive study of two-particle
correlations, such as the HBT correlations of pions, kaons, and protons.
One may investigate also the correlations of non-identical particles, 
this way getting the information about relative distances between
their emission points. All such tasks may be achieved with a complete inclusion of 
resonance decays. 

An extension of the freeze-out geometry of the model to a cylindrically non-symmetric case, 
such as emerging in non-central collisions, would allow for studies of the 
elliptic flow. As already mentioned, an analogous extension to
boost non-invariant geometry would prepare ground for studies of the dependence of observables on the rapidity
variable $y$. Such extensions are planned by the authors in the future.

\section{Installation}
\label{sec:installation}

{\tt THERMINATOR} is distributed in a form of a .tar.gz archive containing the c++ source and the data files.
Running under {\tt ROOT} ver. 4.0x and compiler gcc 3.x has been tested.
In order to run {\tt THERMINATOR} one needs:
\begin{itemize}
\setlength{\itemsep}{-0.1cm}
    \item {\tt c++} compiler
    \item {\tt ROOT} libraries and include files
\end{itemize}
After unpacking the .tar.gz archive in a separate directory, one
compiles {\tt THERMINATOR} by issuing the {\tt make} command in this
directory. 
The executable of {\tt THERMINATOR} - {\tt therm\_events} is
created in the same directory, as well as the {\tt therm\_tree} converting
program. The {\tt SHARE} input files are placed in the {\tt share} subdirectory.

\section{Organization of the {\tt c++} package}
\label{sec:organization}
\subsection{Directory tree}
\label{dir-struct}
\begin{description}
\setlength{\labelwidth}{1.7cm}
\setlength{\itemindent}{0.7cm}
\item [\underline{source code (main directory):}] 
\item [] therm\_events.cxx -- {\tt main program}
\item []ParticleDB.cxx, ParticleDB.h -- {\tt ParticleDB class}
\item []ParticleType.cxx, ParticleType.h -- {\tt ParticleType class}
\item []ReadPar.cxx, ReadPar.h -- {\tt ReadPar class}
\item []Particle.cxx, Particle.h -- {\tt Particle class}
\item []DecayTable.cxx, DecayTable.h -- {\tt DecayTable class}
\item []DecayChannel.cxx, DecayChannel.h -- {\tt DecayChannel class}
\item []Event.cxx, Event.h -- {\tt Event class}
\item []ParticleDecayer.cxx, ParticleDecayer.h -- {\tt ParticleDecayer class}
\item []Integrator.cxx, Integrator.h -- {\tt Integrator class}
\item []Parser.cxx, Parser.h -- {\tt Parser class}
\item []THGlobal.h -- {\tt global helper definitions}
\item []Makefile -- {\tt The makefile}
\item []therm\_tree.C -- {\tt source of the text-to-TTree conversion macro}
\setlength{\labelwidth}{1.7cm}
\setlength{\itemindent}{0.7cm}
\item  [\underline{Data files (share subdirectory):}]
\item  []particles.data --  Particle data file
\item  []decays.data --     Particle decays data file
\item  [\underline{Input files (main directory):}]
\item  therminator.in -- Default input data file     
\end{description}

\section{How to run {\tt THERMINATOR}}

It is assumed that {\tt THERMINATOR} has been successfully compiled and the
{\tt therm\_events} executable is in the user's search path. The user must also know
the location of the {\tt SHARE} input files {\tt particles.data} and {\tt decays.data}.

\subsection{Instructions for the user}

First, the user must examine/edit the {\tt therminator.in} file. If the name of the
file is changed, it then must be supplied as a command-line parameter to the
{\tt therm\_events} program. The meaning of the various parameters is
described in section {\ref{inpfile}}. The input file is also
conveniently commented and contains reasonable default values. A
default input file is Fig. {\ref{inpfex}}. Providing the input file is
in the current directory, one can issue the command:
\begin{verbatim}
therm_events
\end{verbatim}
This runs the program, which  generates the maximum integrand
values, the average multiplicities, and the events themselves. As a
result a text output file with the default name {\tt event.out} is created. Then the user
can issue the command
\begin{verbatim}
therm_tree
\end{verbatim}
and the output file is converted into the {\tt ROOT TFile} {\tt event1.root}
containing the {\tt TTree} named {\tt particles}.
If the user changes the name of the output file from {\tt therm\_events} to {\tt <name>.out}, he must supply this name 
to {\tt therm\_tree} as a command-line parameter. 
The output file from {\tt therm\_tree} will then be called {\tt <name>1.root}.

In case where more that 500 events are requested, the output {\tt ROOT} files are 
divided into {\tt n} batches of maximum 500 events, with the names {\tt <name>n.root}. 

\begin{figure}[tb]
\begin{center}
\includegraphics[width=12cm]{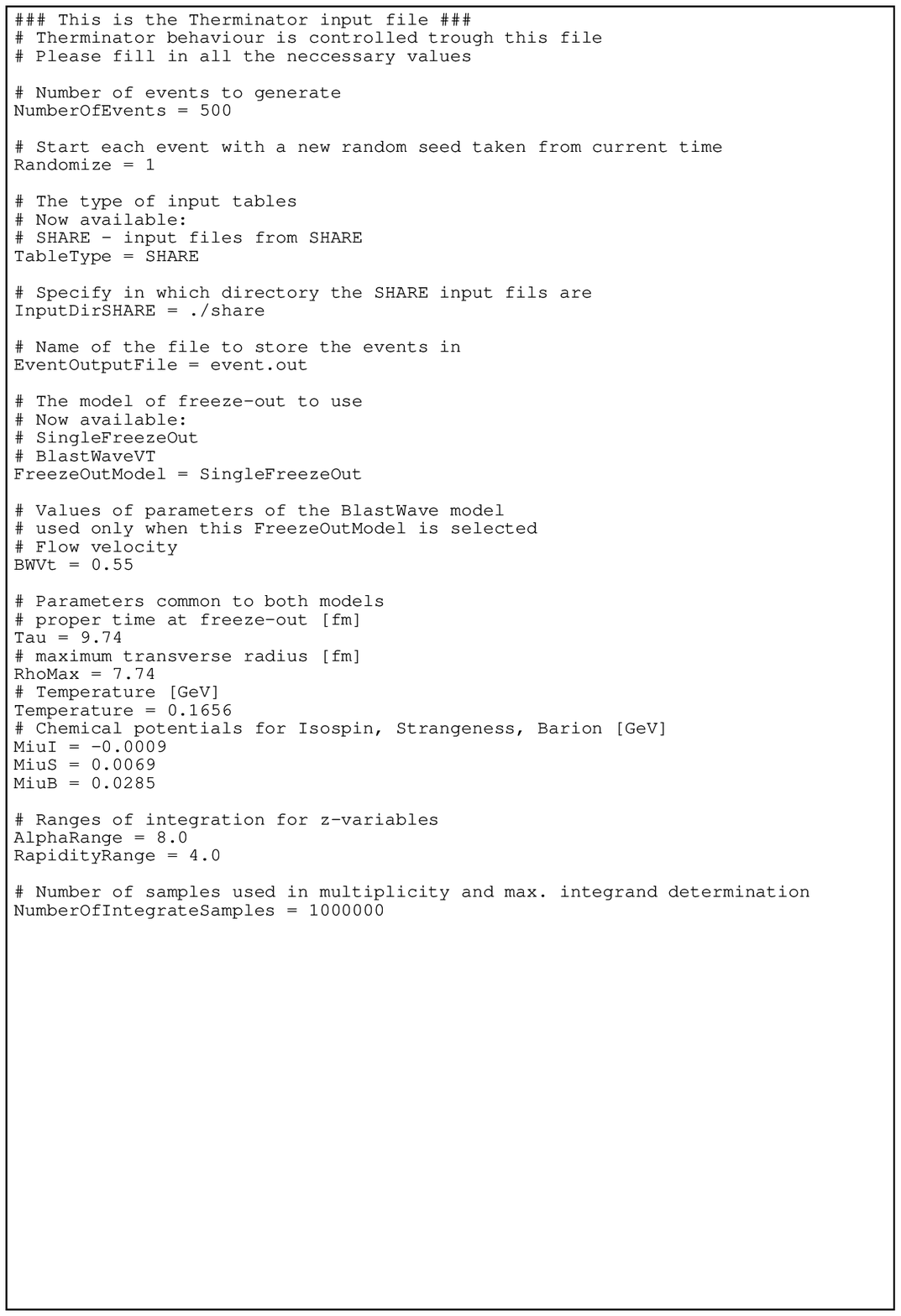}
\end{center}
\caption{An example (default) input parameter file {\tt therminator.in}}
\label{inpfex}
\end{figure}

\subsection{Sample run}

This sample run assumes that the user is in a directory where the default																	
{\tt therminator.in} file shown in Fig. {\ref{inpfex}} is located and
that the {\tt SHARE} input files are in the {\tt share} subdirectory. After
issuing the {\tt therm\_events} and {\tt therm\_tree} commands the
directory should contain the {\tt event.out} ascii output file as well as
the {\tt event1.root} output file. The format of the
{\tt event.out} file is explained in Fig. {\ref{outfex}}. 

\section{Analysis with {\tt ROOT}}

The great advantage of the organization of {\tt THERMINATOR} is the 
compliance to the {\tt ROOT} standard. Having the {\tt event1.root} file, 
it is possible to browse the event tree and generate physical results, even the
involved ones, in a very simple manner. It is assumed that the user knows at least
the basics of {\tt ROOT}. 

The user can start a {\tt ROOT} interactive session: 
\begin{verbatim}
root event1.root
\end{verbatim}
and produce some simple plots from the {\tt ROOT} command line. For example, a
plot similar to the squares plot of Fig. {\ref{ptsp}} can be obtained with the command
(typed in one line) such as
\begin{verbatim}
particles->Draw("sqrt(px*px+py*py)", "(1.0/sqrt(px*px+py*py))*
(pid==211&&fatherpid==211&&abs(0.5*log((e+pz)/(e-pz)))<0.5)")
\end{verbatim}
The contribution from the omega resonance (PDG PID = 223) to the pion
$m_T$-spectra can be viewed with a different command,
\begin{verbatim}
particles->Draw("sqrt(mass*mass+px*px+py*py)",
"(1.0/sqrt(mass*mass+px*px+py*py))*
(pid==211&&fatherpid==223&&abs(0.5*log((e+pz)/(e-pz)))<0.5)")
\end{verbatim}
The correlation between particle's momentum and position (as seen in
Fig.~{\ref{xpx}} can be obtained with
\begin{verbatim}
particles->Draw("px:x>>hxpx(100,-20,20,100,-1.5,1.5)",
"pid==211&&abs(0.5*log((e+pz)/(e-pz)))<0.5")
\end{verbatim}
The particle identification numbers, useful in this interactive 
analysis, can be read from the {\tt particles.data} file.

More elaborated examples, in particular the macros used to generate
the figures of this paper ({\tt figure5.C}, {\em etc.}) are included
in the {\tt THERMINATOR} distribution. To use data from more than one {\tt
eventn.root} file at the same time, use the {\tt Add} method of the
{\tt ROOT} {\tt TChain} class several times. An example of the method
usage in shown in the figure macros.

\begin{figure}[tb]
\begin{center}
\includegraphics[width=14cm]{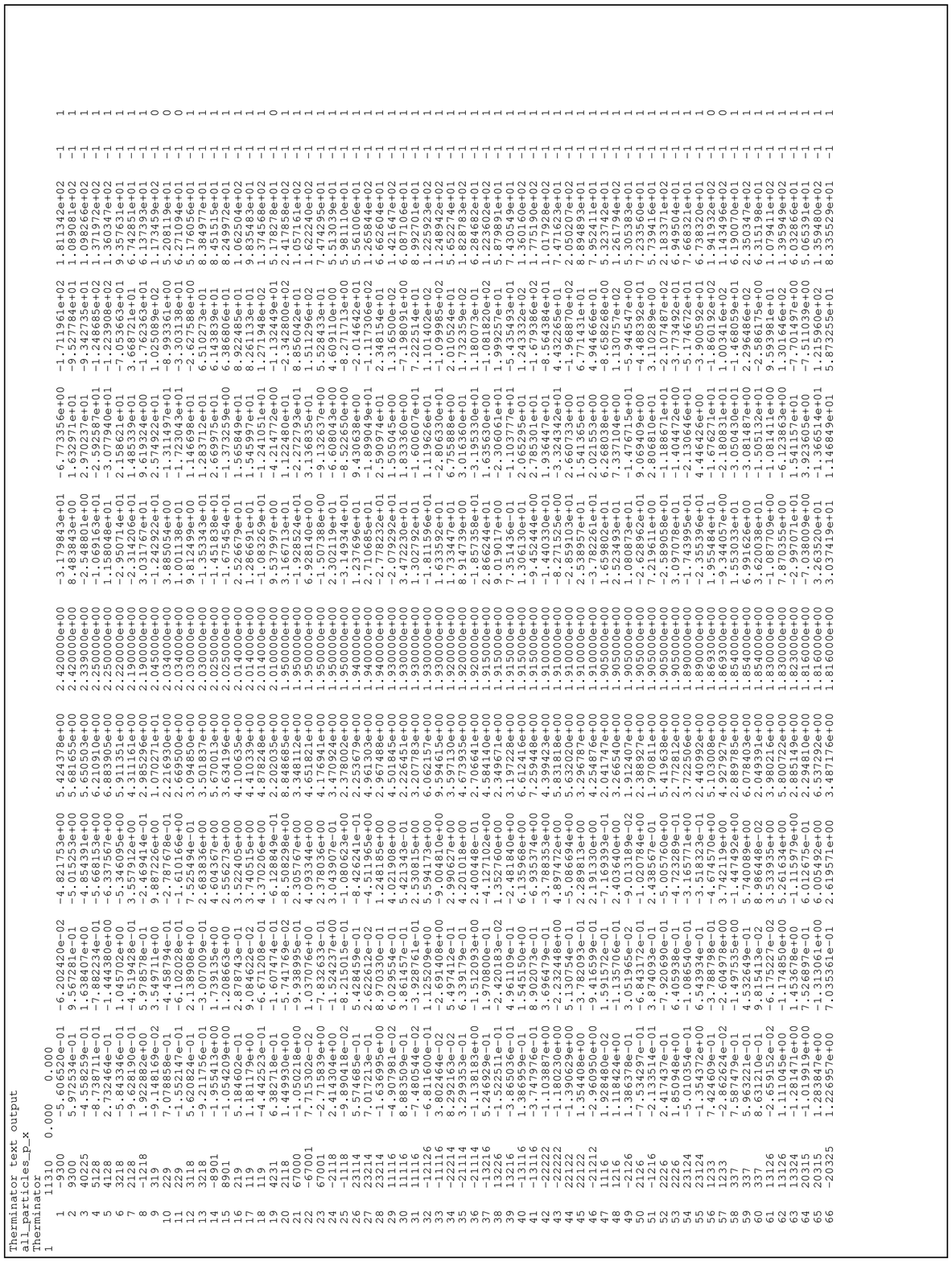}
\end{center}
\caption{An example of the output {\tt event.out}. See Sec. \ref{sect:stor} for details. }
\label{outfex}
\end{figure}

\section{Conclusion}

We hope that {\tt THERMINATOR} will become 
a versatile tool allowing for a simple implementation of the thermal 
approach in analyses of heavy-ion experiments. It may be 
useful for theorists exploring their ideas, as well as 
experimentalists searching for simple and working models which can be
confronted to the data. The results of the code can be directly interfaced 
to {\tt ROOT} -- An Object-Oriented Data Analysis Framework. This way, 
{\tt THERMINATOR} becomes fully compatible with the basic modern software tool
of particle physics. 

New possibilities offered by the program were illustrated in Sec. \ref{sec:new}.
The big advantage offered by {\tt THERMINATOR} is the possibility of exploring
the space-time evolution of the particle system formed in relativistic
heavy-ion collisions. This allows for many interesting physics studies
which should be carried out in the future, such as exploration of various 
freeze-out surfaces, calculation of the HBT correlations with full treatment of resonance decays,  
analysis of the freeze-out criteria, investigation of the role of resonances, 
calculation of balance functions
in various variables, analysis of correlations of non-identical particles, {\em etc}. 
Future extensions of the code should incorporate automatic fitting procedures 
of the model parameters to various observables, such as the transverse-momentum
spectra or HBT radii. Also, the straightforward extension to cylindrically non-symmetric and boost 
non-invariant geometries is planned, which would allow for further interesting 
studies. 

\section*{Acknowledgments}
We thank Jan Pluta for his helpful comments and continuous encouragement expressed
during the realization of this project.

\bibliography{therm10}
\bibliographystyle{h-elsevier3.bst}

\end{document}